\documentstyle[11pt,aaspp4]{article}
\newcommand{\lsim}{$\stackrel{<}{_{\sim}}$}
\newcommand{\gsim}{$\stackrel{>}{_{\sim}}$}
 
\slugcomment{to be published in The Astrophysical Journal, Jan 1, 2002}

\lefthead{Geballe et al.}
\righthead{}

\def\etal{{et al.}} 

\begin{document}
 
\title{Towards Spectral Classification of L and T Dwarfs:  \\
Infrared and Optical Spectroscopy and Analysis}

\author{
T. R. Geballe\altaffilmark{\ref{Gemini}},
G. R. Knapp\altaffilmark{\ref{Princeton}},
S. K. Leggett\altaffilmark{\ref{UKIRT}},
X. Fan\altaffilmark{\ref{IAS}},
D. A. Golimowski\altaffilmark{\ref{JHU}},
S. Anderson\altaffilmark{\ref{Washington}},
J. Brinkmann\altaffilmark{\ref{APO}},
I. Csabai\altaffilmark{\ref{JHU},\ref{Eotvos}},
J. E. Gunn\altaffilmark{\ref{Princeton}},
S. L. Hawley\altaffilmark{\ref{Washington}},
G. Hennessy\altaffilmark{\ref{USNO}},
T. J. Henry\altaffilmark{\ref{JHU}},
G. J. Hill\altaffilmark{\ref{Texas}},
R. B. Hindsley\altaffilmark{\ref{USNO}},
Z. Ivezi\'{c}\altaffilmark{\ref{Princeton}},
R. H. Lupton\altaffilmark{\ref{Princeton}},
A. McDaniel\altaffilmark{\ref{Princeton}},
J. A. Munn\altaffilmark{\ref{Flagstaff}},
V. K. Narayanan\altaffilmark{\ref{Princeton}},
E. Peng\altaffilmark{\ref{JHU}},
J. R. Pier\altaffilmark{\ref{Flagstaff}}, 
C. M. Rockosi\altaffilmark{\ref{Chicago}},
D. P. Schneider\altaffilmark{\ref{PennState}},
J. Allyn Smith\altaffilmark{\ref{Michigan}}
M. A. Strauss\altaffilmark{\ref{Princeton}},
Z. I. Tsvetanov\altaffilmark{\ref{JHU}},
A. Uomoto\altaffilmark{\ref{JHU}},
D. G. York\altaffilmark{\ref{Chicago}}, and
W. Zheng\altaffilmark{\ref{JHU}}
}

\newcounter{address}
\setcounter{address}{1}
\altaffiltext{\theaddress}{Gemini Observatory, 670 North A'ohoku Place,
Hilo, HI 96720
\label{Gemini}}
\addtocounter{address}{1}
\altaffiltext{\theaddress}{Princeton University Observatory, Princeton,
NJ 08544
\label{Princeton}}
\addtocounter{address}{1}
\altaffiltext{\theaddress}{Joint Astronomy Centre, 660 North A'ohoku
Place, Hilo, Hawaii 96720
\label{UKIRT}}
\addtocounter{address}{1}
\altaffiltext{\theaddress}{Institute for Advanced Study, Olden Lane, 
Princeton, NJ 08540 
\label{IAS}}  
\addtocounter{address}{1} 
\altaffiltext{\theaddress}
{Department of Physics and Astronomy, The Johns Hopkins University,
   3701 San Martin Drive, Baltimore, MD 21218, USA
\label{JHU}}
\addtocounter{address}{1}
\altaffiltext{\theaddress}{Apache Point Observatory, P.O. Box 59,
  Sunspot, NM 88349-0059
  \label{APO}}
\addtocounter{address}{1}
\altaffiltext{\theaddress}{Department of Physics of Complex Systems,
E\"otv\"os University,
P\'azm\'any P\'eter s\'et\'any 1/A, Budapest, H-1117, Hungary
\label{Eotvos}}
\addtocounter{address}{1}
\altaffiltext{\theaddress}{University of Washington, Department of
Astronomy, Box 351580, Seattle, WA 98195
\label{Washington}}
\addtocounter{address}{1}
\altaffiltext{\theaddress}{Department of Astronomy, McDonald Observatory,
   University of Texas, Austin, TX~78712.
\label{Texas}}
\addtocounter{address}{1}
\altaffiltext{\theaddress}{U.S. Naval Observatory,
3450 Massachusetts Ave., NW, Washington, DC  20392-5420
\label{USNO}}
\addtocounter{address}{1}
\altaffiltext{\theaddress}{U.S. Naval Observatory, Flagstaff Station,
P.O. Box 1149,
Flagstaff, AZ  86002-1149   
\label{Flagstaff}}
\addtocounter{address}{1}
\altaffiltext{\theaddress}{Department of Astronomy and Astrophysics,
The Pennsylvania State University,
University Park, PA 16802
\label{PennState}}
\addtocounter{address}{1}
\altaffiltext{\theaddress}{University of Michigan, Department of Physics,
500 East University, Ann Arbor, MI 48109
\label{Michigan}}
\addtocounter{address}{1}
\altaffiltext{\theaddress}{University of Chicago, Astronomy \&
Astrophysics
Center, 5640 S. Ellis Ave., Chicago, IL 60637
\label{Chicago}}

%\altaffiltext{\theaddress}{Fermi National Accelerator Laboratory, P.O. 
%Box 500, Batavia, IL 60510
%\label{Fermilab}}
%\addtocounter{address}{1}
%\altaffiltext{\theaddress}{Institute for Cosmic Ray Research,
%University of
%Tokyo, Midori, Tanashi, Tokyo 188-8502, Japan
%\label{CosmicRay}}
%\addtocounter{address}{1}
%\addtocounter{address}{1}

\begin{abstract} We present 0.6--2.5~$\mu$m, R~\gsim~400 spectra of
twenty-seven cool, low luminosity stars and substellar objects. Based on
these and previously published spectra we develop a preliminary spectral
classification system for L and T dwarfs.  For late L and T types the
classification system is based entirely on four spectral indices in the
1-2.5~$\mu$m interval.  Two of these indices are derived from water
absorption bands at 1.15~$\mu$m and 1.4~$\mu$m, the latter of which shows
a smooth increase in depth through the L and T sequences and can be used
to classify both spectral types.  The other two indices make use of
methane absorption features in the H and K bands, with the K band index
also applicable to mid to late L dwarfs. Continuum indices shortward of
1~$\mu$m used by previous authors to classify L dwarfs are found to be
useful only through mid L subclasses. We employ the 1.5~$\mu$m water index
and the 2.2~$\mu$m methane index to complete the L classification through
L9.5 and to link the new system with a modified version of the 2MASS
``Color-d'' index. By correlating the depths of the methane and water
absorption features, we establish a T spectral sequence from types T0 to
T8, based on all four indices, which is a smooth continuation of the L
sequence. We reclassify two 2MASS L8 dwarfs as L9 and L9.5 and identify
one SDSS object as L9. In the proposed system methane absorption appears
in the K band approximately at L8, two subclasses earlier than its
appearance in the H band. The L and T spectral classes are distinguished
by the absence and presence, respectively, of H band methane absorption.

\end{abstract}

\keywords{brown dwarfs; stars: low mass; spectral types; surveys}
 
\section{Introduction}

In the last six years, brown dwarfs, objects of mass too low to sustain
equilibrium hydrogen burning, have gone from being a theoretical concept
with no observed examples to being observationally ubiquitous (Basri
2000). Young brown dwarfs have been discovered in large numbers in star-
forming regions (e.g. Hambly et al. 1999; Pinfield et al. 2000; Luhman et
al. 2000; Lucas and Roche 2000). Objects with masses ranging from just
substellar to those of the Solar System's giant planets have been found
directly or indirectly as companions to nearby stars (Becklin \& Zuckerman
1988; Nakajima et al. 1995; Oppenheimer et al. 1995; Mayor \& Queloz 1997;
Marcy \& Butler 1998, 2000;  Butler et al. 2000;  Marcy, Cochran \& Mayor
2000). In addition, large numbers of isolated brown dwarfs have been found
in the solar neighborhood (Reid \& Hawley 2000; Reid 2001; Kirkpatrick
2001).  These observations have led to controversy regarding the
distinction between brown dwarfs and planets, with some suggesting that
the lower mass limit for brown dwarfs be defined by the ability to burn
deuterium (mass greater than $\sim$13~$M_J$, Burrows et al 1999), and
others preferring the definition that a brown dwarf is any substellar
object that forms via the same process as a star (i.e. not from a
protoplanetary disk).

Brown dwarfs are of interest in their own right. They have compact
structures, low temperatures, and atmospheric constituents that are
markedly different from low mass stars.  The lowest mass brown dwarfs more
closely resemble the gas giant planets than stars, and as such provide
insight into the physical and chemical properties of extrasolar giant
planets that are at present inaccessible for direct study. After an
initial period of deuterium-burning and, for the more massive brown
dwarfs, hydrogen-burning, there is no internal energy source and a brown
dwarf evolves without reaching thermal equilibrium. This leads to age- and
mass-dependent spectral types, in contrast to main sequence stars, and
indeed eventually to effective temperatures lower than that of any stellar
photosphere.

It is generally agreed that the very coolest stars and the warmer brown
dwarfs (1300~K~\lsim~T$_{eff}$~\lsim~2200~K) require a new spectral class,
commonly known as `L' (Mart\'{\i}n et al. 1997; Kirkpatrick et al. 1999,
hereafter K99) and currently known cooler brown dwarfs
(T$_{eff}$~\lsim~1300~K), require the addition of an additional class, `T'
(K99). The L dwarf sequence is characterized by the disappearance of the
red TiO and VO bands from the optical (0.6--1.0~$\mu$m) spectrum, the
increasing dominance at those wavelengths by broad absorption resonance
lines of Na~I and K~I, and strong $\rm H_2O$ absorption bands and
persistent CO overtone bands in the 1-2.5~$\mu$m region (Mart\'{\i}n et
al. 1999, hereafter M99; K99; Kirkpatrick et al. 2000, hereafter K00;
Leggett et al. 2001a). In T dwarfs the CO bands are replaced by stronger
and more extensive absorptions of CH$_{4}$ in the H and K bands, and there
is further strengthening of the water bands (Oppenheimer et al. 1995;
Geballe et al. 1996; Strauss et al. 1999; Burgasser et al. 1999; Cuby et
al. 1999;  Tsvetanov et al. 2000).  The range of L spectral types has been
characterized using optical spectroscopy of objects from the Deep
Near-Infrared Survey, DENIS (Delfosse et al. 1997) and the Two Micron All
Sky Survey, 2MASS (Skrutskie et al. 1997), while first attempts to
describe the range of the `T' spectral sequence are given by Burgasser
(2001), Burgasser, Kirkpatrick, \& Brown (2001a, hereafter B01a), and by
Leggett et al. (2000b, hereafter L00b), whose sample is drawn from the
imaging data of the Sloan Digital Sky Survey, SDSS (York et al. 2000). The
SDSS imaging data allow the discovery of T dwarfs covering a fairly wide
range of spectral characteristics. In pure infrared surveys such as 2MASS,
the degeneracy of the colors of T dwarfs with those of stars (see L00b)
permits only the latest T dwarfs to be selected readily (without the use
of optical photometry).

This paper presents optical (0.6--1.0~$\mu$m) and infrared
(1.0--2.5~$\mu$m) R~\gsim~400 spectroscopy of new L and T dwarfs selected
from the SDSS together with new observations of some previously known M,
L, and T dwarfs. Following the discovery of transition objects with
properties between those of the then known L and T classes (L00b), our
goal has been to obtain sufficient spectra of L and T types over the
entire 0.6-2.5~$\mu$m range and at sufficiently high resolution to allow
the continuous characterization of the L and T dwarf spectral sequences.
Until now the L sequence has been described only on the basis of optical
spectra, and hence it has not been not clear if its full extent has been
observed. In addition there are significant differences in the assignments
of L subclasses for the latest objects (e.g., K99, M99). Using new and
previously published spectra, we have devised a set of infrared flux
indices from which we have defined a complete L spectral sequence, a T
spectral sequence extending to the coolest known brown dwarfs, and a
distinct boundary between the L and T sequences.  One of the new infrared
indices provides a firm link between our L sequence and those proposed by
M99 and K99 on the basis of optical spectral indices.

The L and T sequences we have created are purely empirical and the ranges
of spectral indices describing them are defined so that they progress
smoothly through the known L and T dwarfs. We have made no attempt to
model the spectra or to derive physical quantites such as temperature or
mass. Some observational gaps remain in the T sequence, and it is quite
likely that the sequence will eventually extend to one or more later
subclasses as lower temperature objects are found.

The next two sections describe the selection of objects and the
observations. In Section 4 we discuss various indices for spectral typing
and present the classification system. Section 5 discusses the results,
with emphasis on the definition of the L//T boundary and the link between
infrared and optical classification of L dwarfs. The conclusions are given
in Section 6.  In an adjoining paper, Burgasser et al. (2001b) present an
independently derived spectral sequence for T dwarfs, which bears many
similarities to the one proposed herein. In a companion paper by Leggett
et al. (2001b, hereafter L01b) we provide $\rm ZJHKL'M'$ photometry of
many of objects presented herein and we discuss the behavior of broad-band
colors as a function of spectral class.

\section{The Sample}

Our sample consists of the following groups of objects: (1) sixteen dwarfs
found in new SDSS imaging data, for which we have obtained the first
spectra; (2) two previously reported SDSS dwarfs, one with an incomplete
infrared spectrum and one with no reported infrared spectrum; (3) One
2MASS object (also found in the SDSS) with no reported infrared spectrum,
(4) eight 2MASS L and T dwarfs for which either only very low resolution
(R$\sim$100) 1--2.5~$\mu$m spectra have reported, or there are no reported
1--2.5~$\mu$m spectra and our JHK photometry (L01b) suggested that
spectroscopy would help maximize the range and completeness of spectral
types covered; and (5) fifty-one M, L and T dwarfs for which 1--2.5~$\mu$m
spectra with R~\gsim~400 have been reported previously. See below and
section 3.3 for details of each of these groups. The complete sample is 78
dwarfs covering a range of types from early M to late T and observed at
infrared wavelengths using medium spectral resolution and mostly with the
same telescope and spectrograph. Optical spectra have been obtained for
many of these.

Of the twenty-seven objects whose infrared spectra were observed
specifically for the present work, sixteen are newly discovered in the
imaging data produced by the SDSS. This sample of SDSS objects is not
complete in any way; it was selected to provide as wide as possible range
of spectral properties of L and T dwarfs. The SDSS images the sky in five
filters ($u$, $g$, $r$, $i$, $z$; Fukugita et al. 1996) covering
0.4--1.0~$\mu$m.  The magnitudes measured in these filters and quoted in
the present paper are denoted by $u^*$, $g^*$, $r^*$, $i^*$, and $z^*$,
because the SDSS photometric system is not yet finalized.  The magnitude
scale is on the $AB_{\nu}$ system (J.B. Oke 1969, unpublished), which was
updated to the $AB_{79}$ system by Oke and Gunn (1983) and to $AB_{95}$ by
Fukugita et al. SDSS photometry is modified from the strict logarithmic
magnitude definition to take account of low signal-to-noise ratio, zero,
or even negative, measurements (Lupton, Gunn, \& Szalay 1999). Data
reduction procedures are described by York et al. (2000) and references
therein to the SDSS Project Book, and by Lupton et al. (2001). The
approximate 5$\sigma$ detection limits for a point source observed with a
PSF of full width at half maximum of $1''$ is about 22.5~mag in $r$ and
$i$, and about 20.8~mag in $z$ (York et al. 2000). Some sources were
only detected in $z$ images, but the $i$ band upper limit indicates that
the object is very red.  Nearly all of these $z$-only objects turn out to
be false alarms, usually cosmic rays.  For the present sample, we required
either a 2-sigma detection in the $i$ band and/or a detection by 2MASS
(Skrutskie et al 1997) to avoid wasting observing time on spurious images.

Infrared photometry of the new SDSS dwarfs was obtained at the United
Kingdom Infrared Telescope (UKIRT), and is reported, along with the SDSS
photometry, in Table~1. The photometry of these objects as well as details
of the observations are included in L01b. Note that after their first
mention in a table or the text, the SDSS and 2MASS object names are
abbreviated by the first four digits of the right ascension and first two
digits of declination.

Figure~1 shows finding charts, taken from the SDSS $z$ images, for twelve
of the sixteen new SDSS objects. Charts for SDSS~0107+00, 0236+00,
0423-04, and 2255-00 are given by Schneider et al. (2002). Figure~2
compares the $i^* - z^*$ colors of the above sample with those of a random
sample of 50,000 high-latitude stars extracted from the SDSS imaging data,
demonstrating the extreme red colors of the late M, L and T dwarfs. Note
that L types tend to have $i^* - z^*$ colors in the range 2--3, whereas
for T types the color is usually greater than 3, but that there is
apparently significant overlap.

\section{Spectroscopy}

\subsection{Optical}

We are engaged in follow-up low-dispersion optical spectroscopy of L and T
dwarf candidates selected from the SDSS imaging data, carried out with the
Hobby-Eberly Telescope (HET) and Astrophysical Research Consortium (ARC)
Telescope. The new observations of twelve of the objects considered in the
present paper (eleven SDSS objects and one 2MASS object) are summarized in
Table~2. Among the SDSS objects for which we analyze new optical data in
this paper are SDSS~0837-00 and SDSS~1021-03, whose infrared spectra were
presented by L00b.

The spectra of nine of the objects were obtained using the Double Imaging
Spectrograph, DIS, built by J. Gunn, M. Carr and R. Lupton and mounted on
the ARC 3.5 m telescope at the Apache Point Observatory. The
low-resolution grating with wavelength coverage 5,300--10,000~\AA{} gives
a dispersion of 7.1~\AA{}/pixel and a resolution of 14~\AA{} on the red
side of the spectrograph. The exposure times were 60 minutes. Spectra of
the other three objects were obtained with the Low Resolution Spectrograph
LRS (Hill et al. 1998a,b) at the prime focus of the HET (Ramsey et al.
1998).  The spectra cover the wavelength range 5100 \AA{}~--~9800~\AA{}
with a resolution of 20 \AA{}. The exposure times were 60 minutes (for
details see Schneider et al. 2002).

Spectra obtained at both telescopes were sky-subtracted and
wavelength-calibrated using arc lamps.  Most of the observations were
taken under good, although not photometric, atmospheric conditions.  
Observations of a variety of photometric standard astronomical objects
were used for initial flux calibration The flux scales were adjusted as
described below to provide a smooth joining of the optical and infrared
spectra.

\subsection{New infrared spectra}

As indicated in Table~3, most of the new infrared spectra were obtained at
UKIRT using CGS4 (Mountain et al. 1990). They were obtained in the J, H
and K bands in several observing runs during 2000 and 2001 and, for a
small number of objects, also in the UKIRT Z band (0.85~--~1.05~$\mu$m).
The resolution was about 25~\AA{} in the Z and J bands and 50~\AA{} in the
H and K bands for all of the objects except the bright T dwarf
2MASS~0559-14, for which a spectral resolution of 12.5~\AA{} in the Z and
J bands and 25~\AA{} in the H and K bands was used. Integration times were
5--45 minutes per band per object.

Wavelength calibration was achieved by obtaining near-simultaneous spectra
of arc lamps. Spectra of bright F dwarf stars were recorded immediately
prior to or following each spectrum of a candidate brown dwarf. We
interpolated the spectra of these stars across strong photosheric
features. The spectra of the brown dwarfs were then divided by the
interpolated stellar spectra and the result multiplied by a blackbody
function corresponding to the temperature of the calibration stars and
scaled to the estimated photometric magnitude of the calibration star in
the observed infrared band. These procedures remove the effects of
telluric absorption, yield the correct broad-band spectral shapes of the
target objects, and provide an approximate flux calibration for each band.
More accurate flux-calibration was achieved by scaling the spectra to the
UKIRT ZJHK photometry presented by L01b (cf. also Table 1).

Some of the new spectra were obtained using the Near-Infrared Spectrometer
NIRSPEC (McLean et al. 2000a) on the Keck II telescope on Mauna Kea,
Hawaii, in December 2000. They were measured in several filters covering
the 0.95--2.5~$\mu$m band, with spectral resolving powers of $\sim$~2000
(resolutions of 5--10~\AA{}). Integration times were 10--40 minute per
filter. Data reduction was as described above.  To improve the
signal-to-noise ratios the NIRSPEC spectra were binned to give resolving
powers of $\sim$400.  One R~$\sim$100 Z-band spectrum of one object,
SDSS~0207+00, obtained with NSFCAM at the NASA Infrared Telescope Facility
(IRTF) on September 23, 2000, is also used in this paper.

In addition to objects selected from the SDSS, we obtained CGS4 spectra of
eight 2MASS objects. They are: 2MASS~1523+30 and 2MASS~1632+19, each
classified as L8 by K99, 2MASS~0310+16 and 2MASS~0328+23, each classified
by K00 as L8 on the basis of Keck LRIS spectra; 2MASS~0559-14, a bright,
relatively warm T dwarf whose low resolution infrared spectrum is reported
by Burgasser et al. (2000); and 2MASS~1047+21, 2MASS~1217-03, and
2MASS~1225-27, late-type T dwarfs with low resolution infrared spectra
reported by Burgasser et al. (1999) and B01a.

The new infrared observations are summarized in Table~3.  The optical
observations of most of the objects in this table are listed in Table~2.  
The optical spectra for 2MASS~0028+15, 2MASS~0328+23, 2MASS~1523+30 and
2MASS~1632+19 were taken from K99 and K00.  Optical and infrared spectra
were smoothly conjoined by one of several techniques. In some cases they
were connected using template objects, i.e. relatively bright objects
taken to be representative of the same spectral class. In other cases
optical spectra were scaled to be equal in wavelength ranges of overlap
with the flux-calibrated infrared spectra.  The spectra are displayed in
Fig.~3.  They are ordered according to the spectral class assigned later
in this paper.

\subsection{Previously reported spectra}

In addition to the twenty-seven objects with newly reported spectra listed
in Table~3, we use previously published infrared spectra, mostly obtained
by CGS4 on UKIRT, of seven additional SDSS objects: the L5 dwarf
SDSS~053951.99-005902.0 (Fan et al. 2000, L00b), the T dwarfs
SDSS~0837-00, SDSS~1021-03, and SDSS~125453.90-012247.4 (L00b),
SDSS~1326-00 (Fan et al. 2000), SDSS~1346.45-003150.4 (Tsvetanov et al.
2000), and SDSS~162414.37+002915.6 (Strauss et al. 1999). We also use the
spectra of M and L dwarfs published by Leggett et al. (2000a, 2001a; see
also Kirkpatrick et al. 1995; Ruiz, Allard, \& Leggett 1997; Tinney et al.
1998) and the spectra of four dwarfs obtained by Reid et al. (2001;
hereafter R01). In the case of 2MASS~031059.9+164816 we have substituted
our own spectra in the J and H bands for those obtained by R01.  The
spectra of other three L dwarfs from R01 (2MASSW~J003615.9+182110,
2MASSW~J074642.5+200032, and 2MASSW~J082519.6+211552) are flux-corrected
using the JHK photometry of these objects in L01b.  Finally, we utilize
the spectra of the T dwarfs Gl~229B (data from Geballe et al. 1996 and
Leggett et al. 1999) and Gl~570D (data from Geballe et al. 2001a). The
final sample, after classification, consists of 36 M dwarfs, 25 L dwarfs
and 17 T dwarfs.

\section{Spectral Classification of late L and T Dwarfs}

\subsection{Previous work and current issues}

The recent discoveries of dozens of dwarfs cooler than those of spectral
class M, and the large number of high-quality digitized spectra obtained
for these dwarfs, have made necessary the extension of the traditional
system of stellar spectral classification. Until recently, classification
efforts have concentrated mainly on defining late M dwarfs, the M/L
transition and a wide range of L dwarfs, using optical spectra.  M99,
K99, and K00 discuss the classification of objects in this range of
spectral types and present spectra and classifications of many tens of
objects spanning the stellar-substellar boundary.  They characterize and
define the range of spectral classes M7 to L8 (to L6 in the case of M99)
by (1) the gradual disappearance of the TiO and VO bands; (2) the
increasing depths and breadths of the resonant alkali metal lines and, in
particular, several metal hydrides; (3) the increasing depth of the $\rm
H_2O$ at 9300~\AA{} band; and (4) the increasing steepness of the
0.6--1.0~$\mu$m spectrum.  The broadening of the K~I lines and the even
greater broadening and depth of the Na~I D lines at 5889/5896 \AA{}
continues into the T dwarf spectra (K99, Liebert et al. 2000), removing
much of the optical flux and contributing to the extreme red
optical--infrared colors of these objects.

At 1~$\mu$m and longward, the M and L dwarfs show strong molecular bands
of FeH at 1.00~$\mu$m; H$_{2}$O at 1.35--1.50~$\mu$m, 1.75-2.05~$\mu$m,
and longward of 2.3~$\mu$m; and CO at 2.3--2.5~$\mu$m; as well as
prominent doublets of K~I (Delfosse et al. 1999; McLean et al. 2000b;
Leggett et al. 2000a, 2001a) in the J band. The beginning of the T
sequence is commonly considered to be marked by the appearance of the $\rm
CH_4$ absorptions at H and K bands. Noll et al. (2000) have shown that the
presence of methane is first detectable in the L band via the Q-branch of
its fundamental band near 3.3~$\mu$m in objects classified optically as
about L5, when it is undetectable in the intrinsically weaker shorter
wavelength absorption bands. As the L band is much more difficult to
observe from the ground than the H and K bands, it does not seem advisable
to use the fundamental band for the purpose of classification (L00b,
Geballe et al. 2001b).

For T dwarfs the increasing depths of the methane bands in the H window
cause a reversal of the trend towards redder $J-H$ colors with later
spectral subclass seen in the M and L dwarfs, eventually giving T dwarfs
blue $J-H$ values.  The same trend reversal is seen in $J-K$, although
there the cause appears to be a combination of (1) CH$_{4}$, which absorbs
flux primarily in the long wavelength half of the band; (2) strong
H$_{2}$O bands which depress the K-band on both its long and short
wavelength edges; and (3) H$_{2}$ pressure-induced absorption lines, which
depress the entire K band (Borysow, J$\phi$rgensen, \& Zheng 1997).  Some
atomic features remain present in T dwarfs and can be seen in Fig.~3.
Expanded spectra of the mid-L dwarf SDSS~0107+00 and the mid-T dwarf
2MASS~0559-14 are shown in Fig. 4.

It is generally recognized that the spectral classification of T dwarfs is
best made on the basis of 1-2.5~$\mu$m spectral behavior rather than that
at longer or shorter wavelengths (e.g., B01a). This is because (1) the
deep and highly pressure-broadened optical Na~I and K~I doublets ensure
that almost all of the flux from T dwarfs is emitted longward of 1~$\mu$m,
and (2) for these objects it is easier to obtain high quality spectra at
1-2.5~$\mu$m, where a number of molecular bands are prominent, than at
longer wavelengths, as discussed above. These characteristics also are
true of L subclasses later than about L5, and it can be argued that the
classification of L dwarfs later than about L5 also should be made on the
basis of 1-2.5~$\mu$m spectra.

Tokunaga \& Kobayashi 1999, R01, and Testi et al. (2001)  have defined
sets of indices based largely or entirely on the 1.4~$\mu$m and
1.85~$\mu$m water bands in order to develop L-dwarf classification schemes
that complement existing optical schemes.  R01, who analyze medium
resolution spectra of eighteen dwarfs, show that their entirely water
band-based system is consistent with the 6300--10000~\AA{} system of K99
from spectral types M7 to L8. Testi et al. (2001), who obtained and
analyzed low resolution spectra of 26 L dwarfs, find that R01's indices
are less satisfactory than a set of six indices which they devise. None of
these indices extend the L sequence beyond L8, the limit of K99's optical
classification scheme. B01a have explored methods of classifying T dwarfs,
finding a number of indices that show monotonic and correlated behavior,
as well as some indications of photospheric gravity effects that may
eventually allow the development of a two-dimensional system.

Whatever spectral signatures are used and wherever the shift is made from
indices shortward of 1~$\mu$m to ones longward, making a smooth transition
from optical to infrared classification poses a potential problem, as does
the joining of the L and T sequences. For example, the onset of CH$_{4}$
absorption in the H and K bands, clearly the marker of the beginning of
the T sequence, may not correspond to a natural ``stopping point'' in an L
sequence defined on the basis of other spectral features in other
wavelength bands.  These issues are addressed here, but should be
revisited as more and better quality spectra both shortward and longward
of 1~$\mu$m become available.

\subsection{Indices and spectral subclasses}

We have examined the spectra described in Section 3 with the intent of
developing a set of infrared (1.0--2.5~$\mu$m) spectral indices that are
useful for classifying late L and T dwarfs.  We have also reexamined the
optical ($<$~1~$\mu$m) indices defined by others.  The indices to be used
should be (1) monotonic and discriminating over large ranges of
subclasses, (2) based upon individual spectral features and nearby
continua that are unaffected by the behavior of nearby, unrelated spectral
features, and (3) useful at sites with reasonable atmospheric transparency
and and with intermediate-sized telescopes.

We initially considered indices based upon prominent atomic absorption
lines longward of 1~$\mu$m, but large scatter in the absorption line
strengths for late L subclasses preclude their use. For example, the
scatter in the strength of the K I doublet at 1.25~$\mu$m is easily seen
in Fig.~3 and in the data of R01.  This scatter is not the result of low
quality spectra, but instead may be caused by variations in metallicity,
gravity, and/or rotational velocity. Infrared molecular absorption bands
offer better possibilities. We also have found that broadband infrared
colors (e.g., $J-K$) are not accurate indicators of subclass; reasons for
the large amount of scatter in some of them are discussed by L01b.

Table~4 lists eight candidate indices for classifying L and T dwarfs.
All the indices are flux ratios, defined by

\begin{equation}
index = \frac{\int_{\lambda_1}^{\lambda_2} f_{\lambda}
d\lambda}{\int_{\lambda_3}^{\lambda_4} f_{\lambda} d\lambda}
\end{equation}

\noindent where the wavelength intervals in the numerator and denominator
have the same length. The wavelength intervals of the most useful infrared
indices are shown in Fig. 4. Scripts for calculating all of the candidate
indices are available from SKL on request. Descriptions of the eight
indices follow.

\begin{itemize}

\item {\it Three measures of the continuum slope at 0.7--1.0~$\mu$m,
based on flux ratios of about 0.8~$\mu$m to 0.75~$\mu$m, 1.0~$\mu$m to
0.75~$\mu$m, and 1.0~$\mu$m to 0.9~$\mu$m.} The first of these in Table~4
is the pseudo-continuum (PC3) slope used by M99 and the second is the
``Color-d'' index defined by K99, modified slightly to avoid the FeH band.
The third, based on the continuum slope near 1.0~$\mu$m, is our invention,
with the wavelength ranges set to exclude the TiO, FeH and metal bands.
Both the PC3 and modified Color-d indices show a smooth progression for M
dwarfs and early L dwarfs, but both are affected by the broad K~I
resonance line absorption, which causes the PC3 index to turn over at
about L5 and the modified Color-d index to saturate at about L7. The
1.0~$\mu$m index defined here shows a large scatter for mid-to-late L
types and appears less useful than the other two indices. It may be of
some value in classifying T dwarfs.  In general, all of the red indices
are subject to inaccuracies when applied to substellar objects because of
the very low fluxes at those wavelengths.

\item {\it An index to measure the $\rm H_2O$ band at 1.15~$\mu$m.} The
definition of the 1.2~$\mu$m index, which mainly measures the depth of the
1.15~$\mu$m band, differs from that used by B01a in that the continuum
wavelength range excludes the K~I doublet at 1.169~$\mu$m. The 1.15~$\mu$m
H$_{2}$O band does not appear until late in the L sequence; prior to that
the index mainly reflects the slow reddening of the continuum in the
1.1--1.3~$\mu$m interval. It is only suitable for accurate subtyping of
the T sequence, where the band strengthens rapidly with later subclass. We
compared the 1.2~$\mu$m index with a similar 1.1~$\mu$m water index, which
instead uses the short wavelength continuum peak near 1.08~$\mu$m. The
latter index shows an increase beginning at spectral class L6, near where
the water band first appears. However, there is considerable scatter in
the late L -- early T range, due to the varying width of the continuum
peak (see e.g., Figs. 3b and 3c); because of this it is less effective
than the 1.2~$\mu$m index.

\item {\it An index to measure the $\rm H_2O$ band at 1.5~$\mu$m.} The
1.5~$\mu$m index defined here measures the slope of the long wavelength
side of the 1.4~$\mu$m water band.  Its definition is slightly different
from that used by R01; the shorter wavelength range is closer to the
center of the 1.4~$\mu$m feature (but still accessible to ground-based
telescopes), and the longer wavelength range better matches the location
of the continuum peak as it appears in T dwarfs (see Fig. 4b). It shows a
steady increase through both the L and T sequences.

\item {\it An index to measure the $\rm H_2O$ band at 1.9~$\mu$m.}
Previous attempts to characterize the 1.9~$\mu$m H$_{2}$O absorption band
in L dwarfs have produced mixed results (Tokunaga \& Kobayashi 1999; M99;
R01; Testi et al. 2001). We have defined a 2.0~$\mu$m index that measures
the redward slope of the 1.9~$\mu$m band while avoiding the strong
telluric CO$_{2}$ absorptions centered at 2.01~$\mu$m and 2.06~$\mu$m.  
Thus, our index resembles that of Testi et al. (2001) but spans a shorter
wavelength range than that of R01. Like Testi et al., we find that this
index suffers more scatter than shorter wavelength $\rm H_2O$ indices.
Also, the range of this index is more compressed than those of the other
water indices, The reason for the relative insensitivity of the 2.0~$\mu$m
index to spectral subclass is not clear. It is probably not due to
continuum emission from dust (Allard et al. 2001; Marley \& Ackerman 2001;
Burrows et al. 2001, Tsuji 2001), which would dilute the strength of the
band, because a similar effect is not seen in the nearby 2.2~$\mu$m
methane band. We conclude that the 2.0~$\mu$m index is not a sufficiently
accurate indicator of L or T subclass.

\item {\it Indices for the CH$_{4}$ bands at 1.6~$\mu$m and 2.2~$\mu$m.}
The definition of the H band methane index differs from that used by B01a;
the wavelength range used here includes both the 1.63~$\mu$m and
1.67$\mu$m absorption maxima seen in Fig.~4b. The index is flat through
the L sequence but increases monotonically through the T sequence.  The
2.2~$\mu$m index differs from those at similar wavelengths employed by R01
and B01a, and is specifically designed to detect the P branch of the
$\nu_2+\nu_3$ methane band, just longward of the Q branch which produces a
narrow absorption feature at 2.20~$\mu$m in T dwarfs (Fig. 4b). The
$\nu_3+\nu_4$ and $\nu_1+\nu_4$ methane bands centered at 2.32~$\mu$m and
2.37~$\mu$m are somewhat stronger than this band.  However, because of the
complexity of the spectra of early T dwarfs at 2.3--2.5~$\mu$m,
with CO, CH$_{4}$, and H$_{2}$O all absorbing significantly, the first
clear sign of methane in the K band is the appearance of the 2.2~$\mu$m
band. Although this band of methane does not become obvious until near the
L/T boundary, the index we have defined may be of some use for classifying
dwarfs as early as L3. For late L subclasses the index is clearly already
increasing with spectral subclass, recording an increasing depression of
the long wavelength side of the K band, and there is some tendency for
this to be occurring at mid L as well.

\end{itemize}

In summary, we find four indices in the 1.0--2.5~$\mu$m interval which are
monotonic over sufficiently wide portions of the L/T sequence that they
are suitable for accurately classifying L and T dwarfs.  Two of the new
indices, at 1.2~$\mu$m and 1.5~$\mu$m, are associated with water
absorptions and two, at 1.6~$\mu$m and 2.2~$\mu$m, are associated with
methane absorptions. In T dwarfs, all but the H$_{2}$O 1.5~$\mu$m index
measure the actual depths of the associated bands; in the L dwarfs the
useful indices measure slopes on the wings of bands (see Fig.~4).  Very
importantly, the 1.5~$\mu$m water band index is suitable for
classification across the entire L--T sequence and can be linked to the
optical continuum indices PC3 and Color-d, which are already being used
for classifying early to mid L types. Although it has a smaller range of
usefulness in the L sequence, the 2.2~$\mu$m index can be used along with
the 1.5~$\mu$m index to smoothly link the L and T classifications.

We can now use these infrared indices as well as the PC3 and modified
Color-d indices to define numerical ranges for spectral subclasses.  This
is done in Table~5. In defining the subclasses our intent is: (1) to make
the numerical values of the infrared and optical indices in the L0--L5
range generally consistent with existing classifications of M99 and K99 in
that range, where the two optical systems agree; (2) to define a sensible
L/T boundary and extend the L sequence smoothly to it; and (3) to define
a T sequence, which continues the smooth progression of the indices that
are used both for L and T classification and which leaves some room in a
system of ten subclasses for as yet undiscovered T dwarfs that are later
than any of those currently observed.

\subsection{Results}

Table~4 lists the spectral indices and derived classifications for all
dwarfs in our sample. For each index listed in Table~4, we determined a
classification to $\pm$0.5 subclass based on the definitions in Table~5.
We assign a ``.5'' subclass if the difference of the value of the index
and the value at the boundary between subclasses is less than 25\% of the
range of the index for that subclass, as defined in Table~5. In the final
column of the table we have assigned spectral classifications to all L and
T objects.  These are based on the unweighted averages of the
classifications from all six usable optical and infrared indices for L5
and earlier, and on the averages of only the four infrared indices for
later objects. The classifications of the M dwarfs in Table~4 are obtained
from the literature, except for SDSS~2255-00, which we have classified to
be consistent with other late type M dwarfs using the infrared water
bands.

Figure~5 contains plots of the observed infrared indices versus final
assigned spectral class. A large majority of the data points for each
index fall within the defined range for that index, demonstrating the
accuracy of the technique. In most cases the L and T classifications are
consistent within the overall classification to within about one-half of a
subclass and the final classification may be considered accurate to that
amount. Larger uncertainties are noted in the last column of Table~4 and
several of these cases are discussed in section 5.1.

\section{Discussion}

\subsection{Comparison with previous classifications and internal
consistency}

Our classifications of L0--L5 dwarfs are in good agreement with those of
M99, K99, and K00 and the new infrared indices and the 0.6-1.0~$\mu$m
continuum indices complement one other in this range of subclasses.
Clearly, optical indices should be weighted heavily when classifying such
objects. For later L dwarfs, however, the 1.5~$\mu$m and 2.2~$\mu$m
indices yield additional information and are more accurate classification
tools than the 0.6--1.0~$\mu$m continuum indices.  As shown in Table~5 we
are unable to define a classification beyond L4 for the PC3 index and
beyond L6 for the Color-d index. The dwarfs for which our classifications
differ significantly from previously published ones are 2MASS~0028+15,
where we obtain L3 whereas K00 find L4.5; 2MASS~0825+21, where we find L6
but K00 and R01 find L7.5; 2MASS~0310+16 and 2MASS~0328+23, which we
classify as L9 and L9.5, respectively, whereas K00 obtain L8 for both; and
2MASS~1632+19, which we find to be L7.5, but M99 classify L6. Most of
these are late-type L dwarfs, where we expect larger discrepancies.

We found that the latest L subclasses described both by M99 and K99
correspond fairly closely to the boundary between L and T classes (i.e.,
to the onset of observable methane absorption in the H and K bands). We
also determined that a continued smooth progression of the 1.5~$\mu$m
water band index to this boundary requires several additional L subclasses
beyond L5, and hence matches the K99 classification system (L0--L8) better
than that of M99 (L0--L6). Therefore, our scheme is loosely correlated
with the remainder of the K99 L sequence, with values of both the
1.5~$\mu$m and 2.2~$\mu$m indices that roughly match the K99
classifications. However, as discussed above, we found a number of
instances in which the values of the two infrared indices implied a
different L classification than that provided by K99. In addition, as
discussed in section 5.2 below, we define an L9 subclass between L8 and
T0.  If generally adopted, this might require a reworking of the optical
classification systems for the later L subclasses.

For L types where both the 1.5~$\mu$m and 2.2$\mu$m indices are measured,
the agreement between them is one subclass or better for 15 of the 23
cases (see Table~4 and Fig.~6). Agreement generally improves with later
subclass. There are three instances (GD~165B, DENIS~0205-11, and
DENIS~1228-15) for which the two indices give widely discrepant results.
The final uncertainties in those cases are $\pm$2 subclasses (see
Table~4); however, in each case our final assignment of spectral class is
consistent with K00 and R01. Each of these dwarfs as well as 2MASS~0345+25
is in a closely spaced multiple system (Becklin \& Zuckerman 1988;
Mart\'{\i}n, Brandner, \& Basri 1999; Koerner et al. 1999; Reid et al.
1999); perhaps they have additional unseen components which lead to the
discrepancies.  They are not plotted in Fig.~6.

No detailed classification system for T dwarfs has been proposed
previously, although some possibilities have been explored by B01a.  
Burgasser et al. (2001b) are also suggesting a scheme for T dwarfs, which
gives similar results as our system. The latest T dwarfs that we
classified are 2MASS~1217-03 and Gl~570D, both T8. The former appears
marginally earlier in three of the four infrared indices. The next latest
object that we have classified is the T6.5 dwarf 2MASS 1047+21.  We have
not yet found examples of T5 or T7 dwarfs.

For the T dwarfs the agreement between the four indices we have used for
classification is generally better than the agreement for the indices used
for the L dwarfs (see Table~4 and Fig.~6). This is because the indices
have much wider ranges of values through the T sequence than through the L
sequence. A single index usually agrees with the final T spectral
classification to $\pm$0.5 or less. Accuracy of classification increases
with later spectral type. Interestingly, the T dwarf with the most
uncertain classification in our system is the ``prototype,'' Gl~229B,
assigned T6. Its deviations, which are easily seen in Fig.~5 (the most
discrepant points at T6), probably are not intrinsic to Gl~229B, but
rather are related to the difficulty in correcting its spectrum for stray
and diffracted light from the nearby primary, which is roughly 10,000
times brighter in the $J$, $H$, and $K$ bands.  Both Gl~229B and
SDSS~1624+00 are classified T6, but we note that both Liebert et al.  
(2000) and Nakajima et al. (2000) find that Gl~229B is cooler, based on
the widths and strengths of atomic lines and its slightly deeper
1.6~$\mu$m methane band, from which one might infer that in a different or
finer system it might be assigned a later subclass than SDSS~1624+00.
Taking our spectra at face value, we find that Gl~229B is later than
SDSS~1624+00 in its 1.6~$\mu$m methane index, but earlier in its
1.5~$\mu$m water and 2.2~$\mu$m methane indices.

\subsection{The L/T boundary and the T sequence}

The numerous spectra of late-L dwarfs and early-T dwarfs permit us to
examine in detail the onset of methane absorption in brown dwarf spectra
and to propose a boundary between the L and T classes. Figure~7 shows a
representative selection of late L and early T spectra in the
1.45-2.50~$\mu$m region, where the critical absorption features and
indices are situated. Close examination of these spectra shows that the
2.20~$\mu$m methane absorption is first noticeable at spectral subclass
L8. The absorption is quite subtle at this subclass, appearing as a slight
inflection near 2.20~$\mu$m. At somewhat later subclasses a flux minimum
is seen at that wavelength. The index measuring this feature also measures
the slope between 2.1$~\mu$m and 2.25~$\mu$m changing steadily from mid to
late L. H$_{2}$ collision-induced absorption is expected to be fairly flat
across the K band (Borysow et al. 1997), due to the contributions by
several H$_{2}$ transitions at different wavelengths in the band and the
breadths of each absorption profile, and probably is not responsible for
this change. The effect of the 2.20~$\mu$m CH$_{4}$ absorption is to
increase the value of the index; by T3 the CH$_{4}$ absorption is
very strong.

In contrast, the methane bands at 1.6--1.7~$\mu$m are not seen at
subclasses L8 and L9 in our system (see Figs. 4b and 7). The spectra of
2MASS~1523+30 and SDSS~0830+48, which we classify as L8 and L9,
respectively, are flat from 1.60~$\mu$m to 1.69~$\mu$m, as are spectra of
earlier L subclasses. The two absorption features at 1.67~$\mu$m (due to
the 2$\nu_{3}$ Q branch) and 1.63~$\mu$m are first clearly seen at T0, two
subclasses later than where the 2.2~$\mu$m ($\nu_{2}+\nu_{3}$) band first
becomes apparent. Unlike the K band methane index, the index measuring the
1.6~$\mu$m methane bands is only useful for distinguishing T subclasses.

Thus the boundary between L and T classifications in our proposed system
corresponds to the appearance of methane absorption in the H band; i.e.,
to its earliest appearance in {\it both} the H and K bands.  It is this
choice of boundary and the maintainance of smoothly varying infrared
indices through it that have led us to define an L9 subclass. The proposed
L/T boundary is preferred over one defined as the onset of CH$_{4}$
absorption in the K band, for several reasons: (1) the noise levels at
shorter infrared wavelengths are generally lower than at longer
wavelengths; (2) the continuum flux levels of brown dwarfs on the L/T
boundary are higher at 1.6~$\mu$m than at 2.2~$\mu$m; and (3) the (later)
first appearance of methane in the H band creates a more prominent and
obvious spectral feature than the (earlier) appearance in the K band.

Figure~8 shows the spectra of T type dwarfs from T0 through T8, with
adjacent spectra spaced by one subclass (except for the gaps between T4.5
and T6 and between T6 and T8). The spectra show a smooth progression
through the sequence, as was the intention when the ranges of indices were
designed for each subclass. Note that the 1.5~$\mu$m index measures the
slope of the shoulder of the 1.4~$\mu$m water band, whereas the other
three indices measure the actual depths of the associated features (see
Fig.~4). However, it is apparent from Fig.~8, and not surprising, that
increasing slope of the shoulders of the 1.4~$\mu$m water band also
corresponds to a steady deepening of that band with later subclass.

\subsection{Linking the infrared and optical indices}

Figure~9 shows the modified Color-d optical index plotted against the
1.5~$\mu$m index for all isolated L dwarfs in which both indices were
measured. The Color-d index saturates above L6, as illustrated in the
figure. For 7 of 8 of the objects in the L0--L6 range the agreement
between the two indices is better than one subclass, and in half it is
better than one-half subclass. The discrepant object is 2MASS~0036+18,
whose Color-d classification is L2.5 and the whose 1.5~$\mu$m and
2.2~$\mu$m classifications are L4.5 and L5, respectively. (Note that K00
have classified this object as L3.5 and that our overall classification is
L4.) The discrepant case notwithstanding, the good correlation of these
indices in the L0-L6 range permits a graceful transition between the
optical and infrared classification schemes.  More data and further
comparison of the 1.5~$\mu$m index with other optical indices are needed
to strengthen this link between the two classification schemes.

\section{Conclusions}

We have developed a system for classifying L and T dwarfs, based largely
on four newly defined spectral indices in the 1.0-2.5~$\mu$m range, and
have used it to classify objects from L0 to T8.  Two new indices we use
are associated with the 1.15 and 1.4~$\mu$m water bands and two are
associated with the 1.65 and 2.2~$\mu$m methane bands. At L0--L6 the new
classification system gives consistent results with the optical systems of
M99 and K99, and it appears that they can be accurately aligned using the
1.5~$\mu$m water and modified Color-d (and possibly other optical)
indices.  For later L types and all T types, the new indices give
satisfactory and self-consistent results.  An L classification based on
either the 1.5~$\mu$m or 2.2~$\mu$m index is usually accurate to within
$\pm$1 subclass, and a T classification using any one of the four infrared
indices is usually accurate to within $\pm$0.5 subclass. The boundary
between the L and T classes has been thoroughly investigated.  We define
the boundary to be the appearance of methane absorption in the H band,
which occurs approximately two spectral subclasses after its appearance in
the K band.

Spectra of additional L and T dwarfs are needed to refine the definitions
of the subclasses. Nevertheless, as defined here the L sequence is
complete, extending from L0 to L9.5 and smoothly connecting to the T
sequence. The same cannot be said of the T sequence, as objects later than
T8, yet to be observed, should exist. Observing the full extent of the T
sequence and determining whether the present classification system is
acceptable or requires compression, or a more radical alteration, needs
the detection and spectral study of cooler and presumably fainter dwarfs
than currently found. The range of effective temperatures of T0-T8 dwarfs
is 800--1300~K (Geballe et al. 2001a; L01b). Radical changes in the
1.0--2.5$~\mu$m spectrum should not occur until temperatures are low
enough that water vapor condenses, probably at T$_{eff}$~$\sim$~500~K
(Marley \& Ackerman 2001).  Hence, compression of the classification
system defined here is a possibility.  Finally, clear understandings of
the effects of varying gravity, rotation, dust, and metal abundance, and
incorporating them into a multi-dimensional classification system will
require higher precision and probably higher resolution spectra than
reported here, together with accurate model atmospheres and spectra.

\acknowledgements

We thank the 2MASS group (A. J. Burgasser, J. D. Kirkpatrick, M. E. Brown,
I. N. Reid and colleagues) for many valuable discussions on spectral
typing, for providing data before publication, and for communicating their
spectral sequence for L and T dwarfs before publication.This shared
information allowed us to make a consistent first set of T dwarf spectral
classifications prior to more detailed analysis. GRK is grateful for
support from Princeton University via an Old Dominion Fellowship and from
NASA via grants NAG-6734 and NAG5-8083. XF acknowledges supports from NSF
grant PHY-0070928 and a Frank and Peggy Taplin Fellowship. DAG
acknowledges support from the Center for Astrophysical Sciences at Johns
Hopkins University. DPS acknowledges support from NSF grant (NSF99-00703).

The Sloan Digital Sky Survey (SDSS) is a joint project of the University
of Chicago, Fermilab, the Institute for Advanced Study, the Japan
Participation Group, the Johns Hopkins University, the Max Planck
Institute for Astronomy (MPIA), the Max Planck Institute for Astrophysics
(MPA), New Mexico State University, Princeton University, the United
States Naval Observatory, and the University of Washington.  Apache Point
Observatory, site of the SDSS telescopes, is operated by the Astrophysical
Research Consortium (ARC). Funding for the project has been provided by
the Alfred P. Sloan Foundation, the National Science Foundation, the U.S.
Department of Energy, the Japanese Monbukagakusho, and the Max Planck
Society.  The SDSS web site is http://www.sdss.org/

UKIRT is operated by the Joint Astronomy Centre on behalf of the UK
Particle Physics and Astronomy Research Council.  Some of the observations
made there were done through the UKIRT Service Program. We thank the staff
of UKIRT for its expert help. We also are grateful to the staff of the
W.M. Keck Observatory, which is operated as a scientific partnership among
the California Institute of Technology, the University of California and
the National Aeronautics and Space Administration and was made possible by
the generous financial support of the W.M. Keck Foundation. The
Hobby-Eberly Telescope (HET) is a joint project of the University of Texas
at Austin, the Pennsylvania State University, Stanford University,
Ludwig-Maximillians-Universit\"at M\"unchen, and
Georg-August-Universit\"at G\"ottingen.  The HET is named in honor of its
principal benefactors, William P. Hobby and Robert E. Eberly.

\newpage 
\begin{figure}
\figurenum{1}
\epsscale{.7}
%\epsfysize=400pt
%\epsfbox{/home/tgeballe/text/ltclass/fig1x.ps}
\plotone{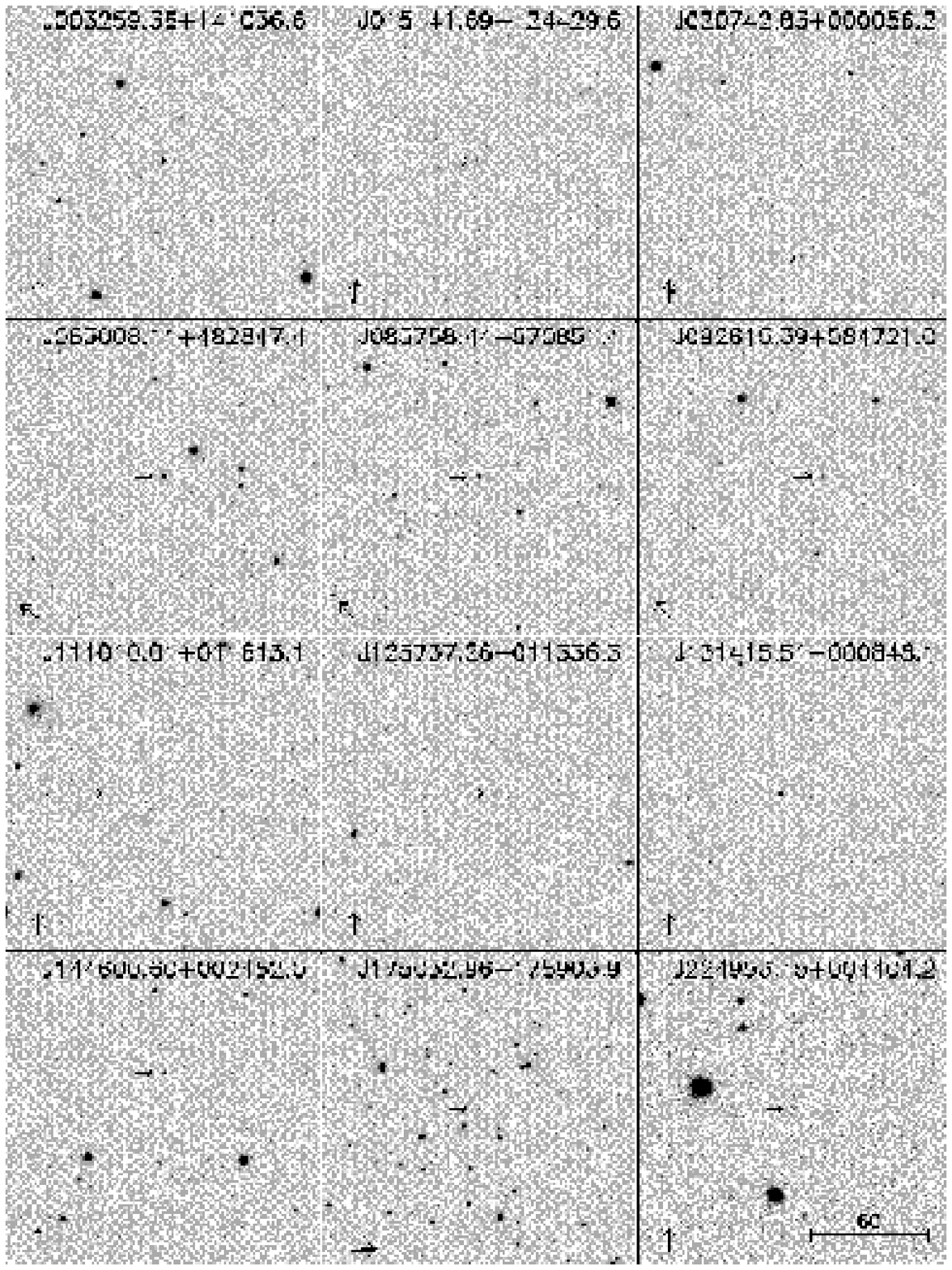}
\caption{$z$ finding charts for twelve new SDSS dwarfs. North is
denoted by the arrow at bottom left of each chart; east is
counterclockwise from north.}
\end{figure}

\newpage

\begin{figure}
\figurenum{2}
\epsscale{.8}
\plotone{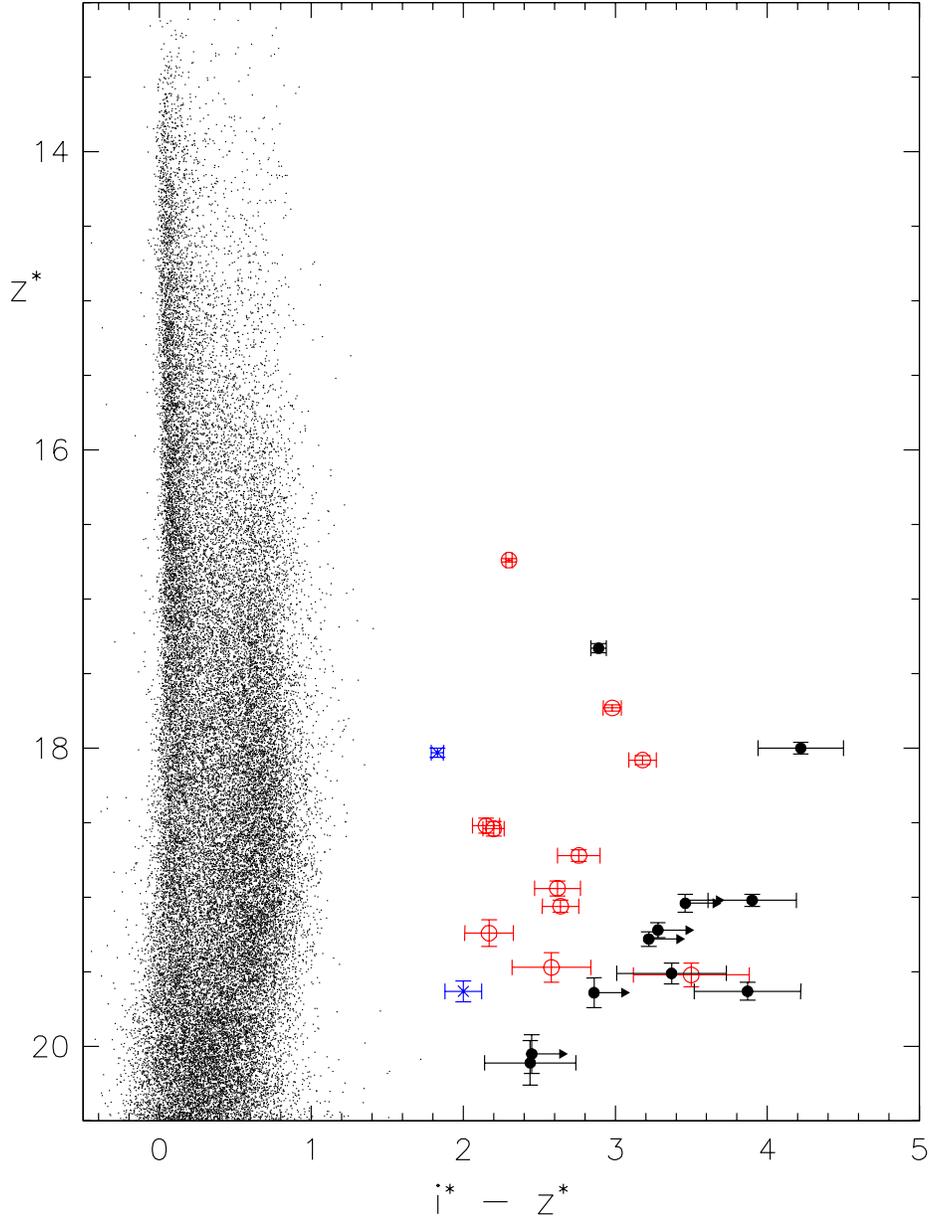}
\caption{Color magnitude diagram, $z^*$ vs. $i^*-z^*$, for
low mass stars and substars selected from the SDSS imaging data and
discussed in the present paper.  For comparison, the data for a
random sample of 50,000 high-latitude stars selected from the SDSS
imaging data are shown - the objects in this sample were selected to
be detected and point-like separately in the SDSS $g$, $r$
and $i$ band data.  The symbols represent the spectral classes
assigned in this paper.  Crosses: M.  Open: L. Filled: T.  The
lower limits to the $i^*-z^*$ colors are calculated assuming the
5$\sigma$ limit of 22.5 magnitudes at $i$ band.  The error
bars are 1$\sigma$, and do not contain a contribution for
systematic zero-point error.}
\end{figure}
\newpage

\begin{figure}
\figurenum{3a}
\epsscale{.9}
\plotone{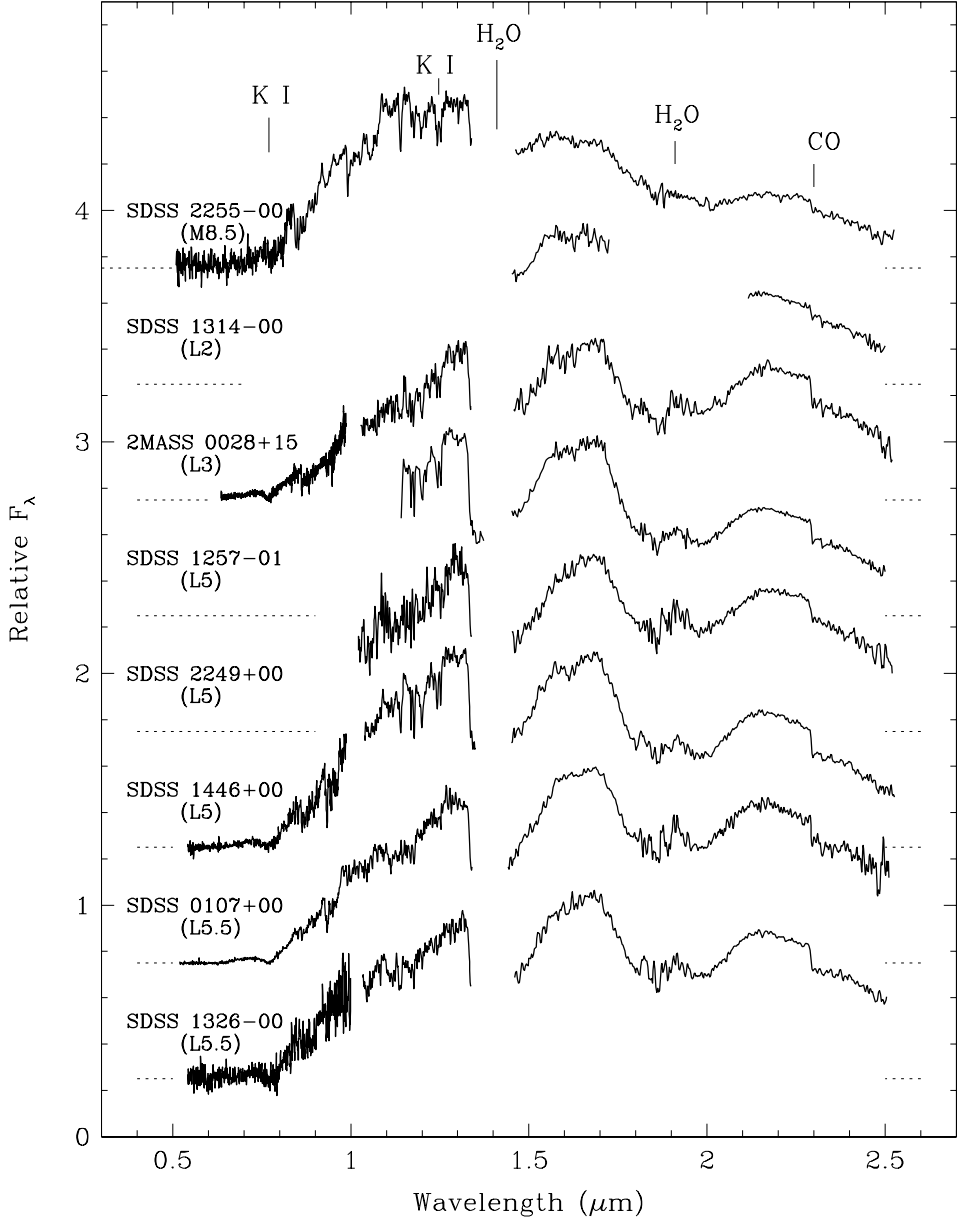}
\caption{New spectra of twenty-seven M, L, and T dwarfs. The
SDSS or 2MASS identifications are each given above the dashed line which
corresponds to the zero flux level for each spectrum. The complete
catalogue names are given in Tables 1-3. The locations of significant
spectral features are indicated at the top of each figure. Classifications
are from this paper. Local noise can be judged by point-to-point
variations.  M9 through L5.5.}
\end{figure}

\newpage

\begin{figure}
\figurenum{3b}
\epsscale{.9}
\plotone{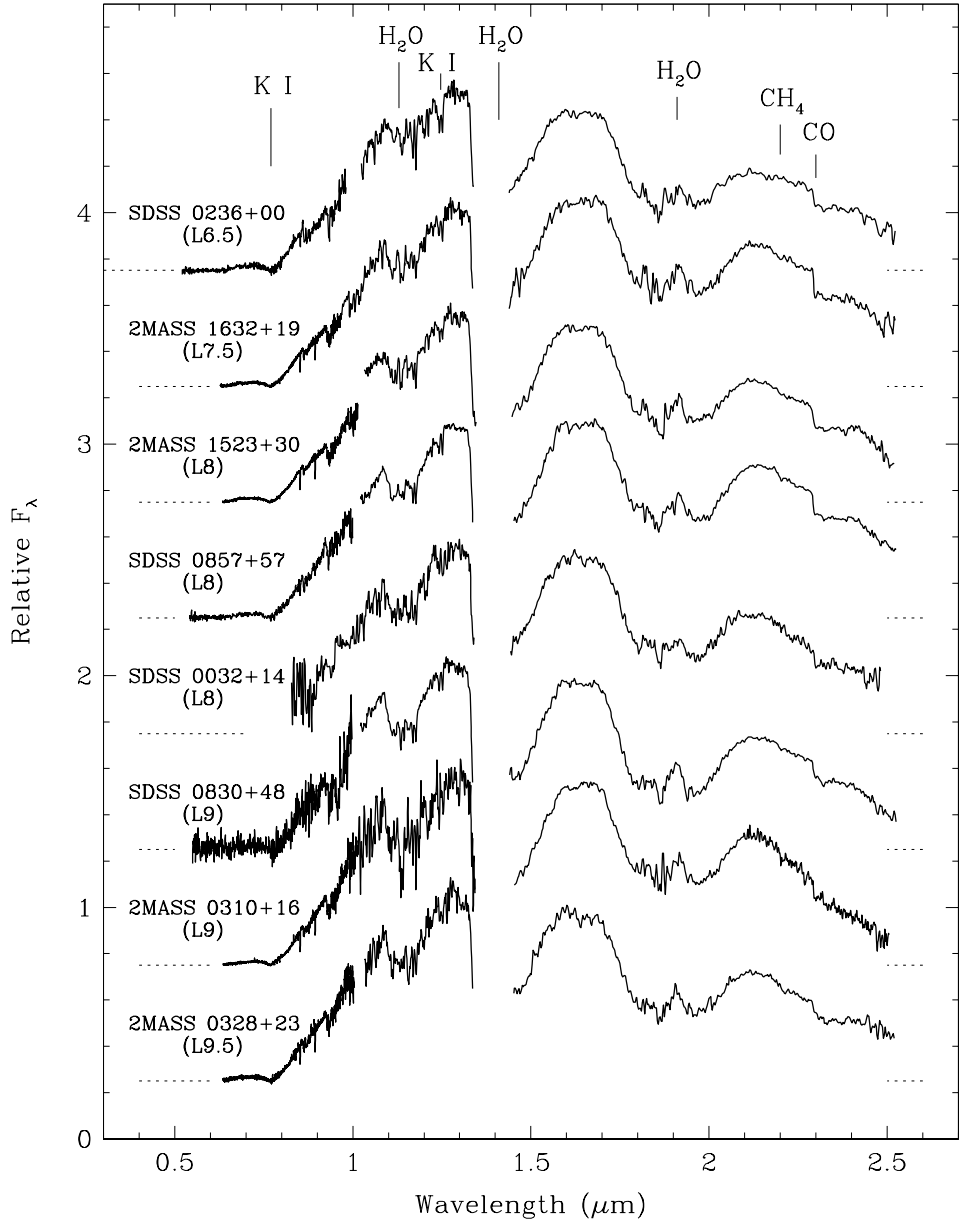}
\caption{L6.5 through L9. 2MASS~0310+16 includes our data (J and H), 
optical data from K00, and infrared data from R01 (K).}
\end{figure} 

\newpage

\begin{figure}
\figurenum{3c}
\epsscale{.9}
\plotone{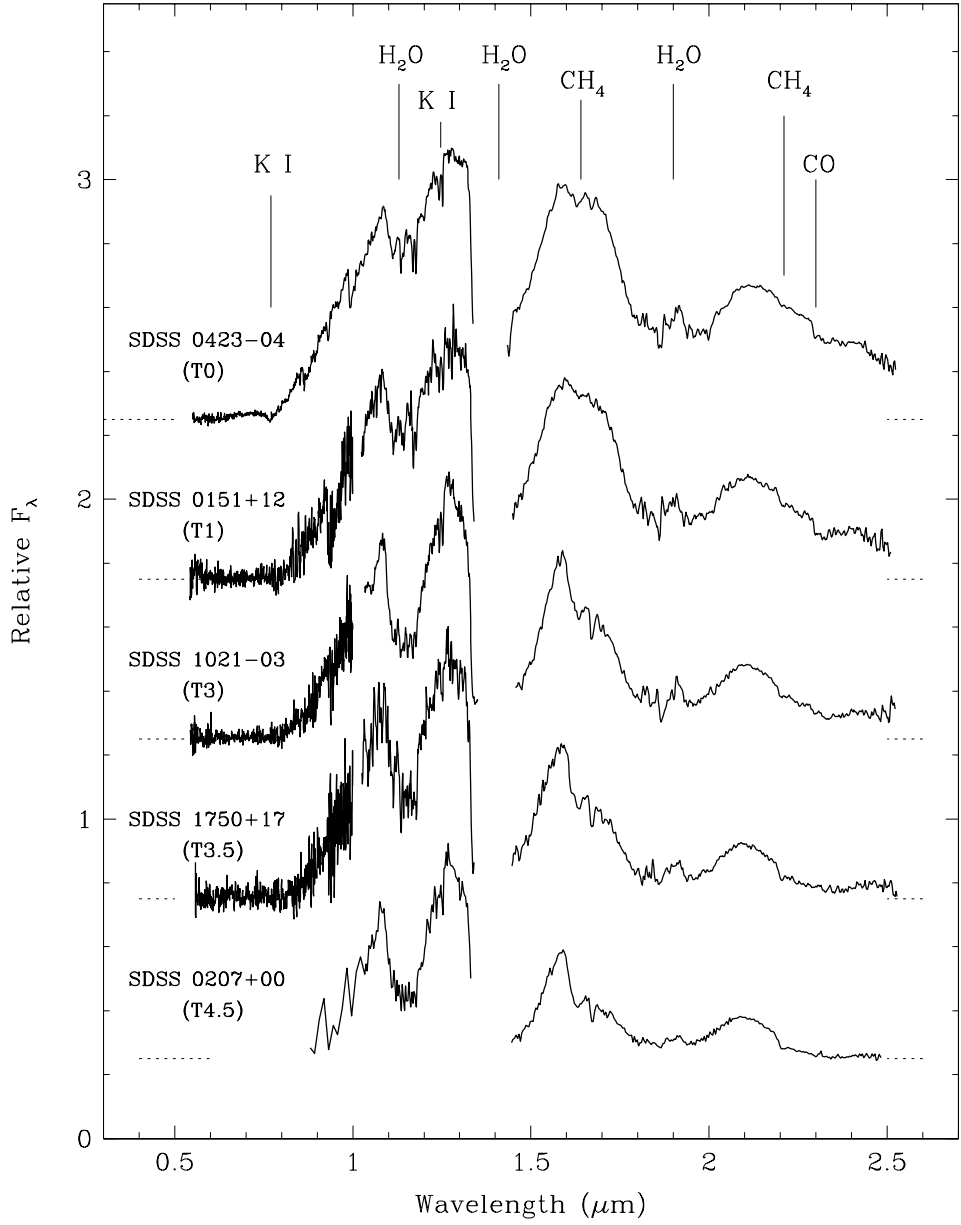}
\caption{T0 through T4.5.}
\end{figure}

\newpage

\begin{figure}
\figurenum{3d}
\epsscale{.9}
\plotone{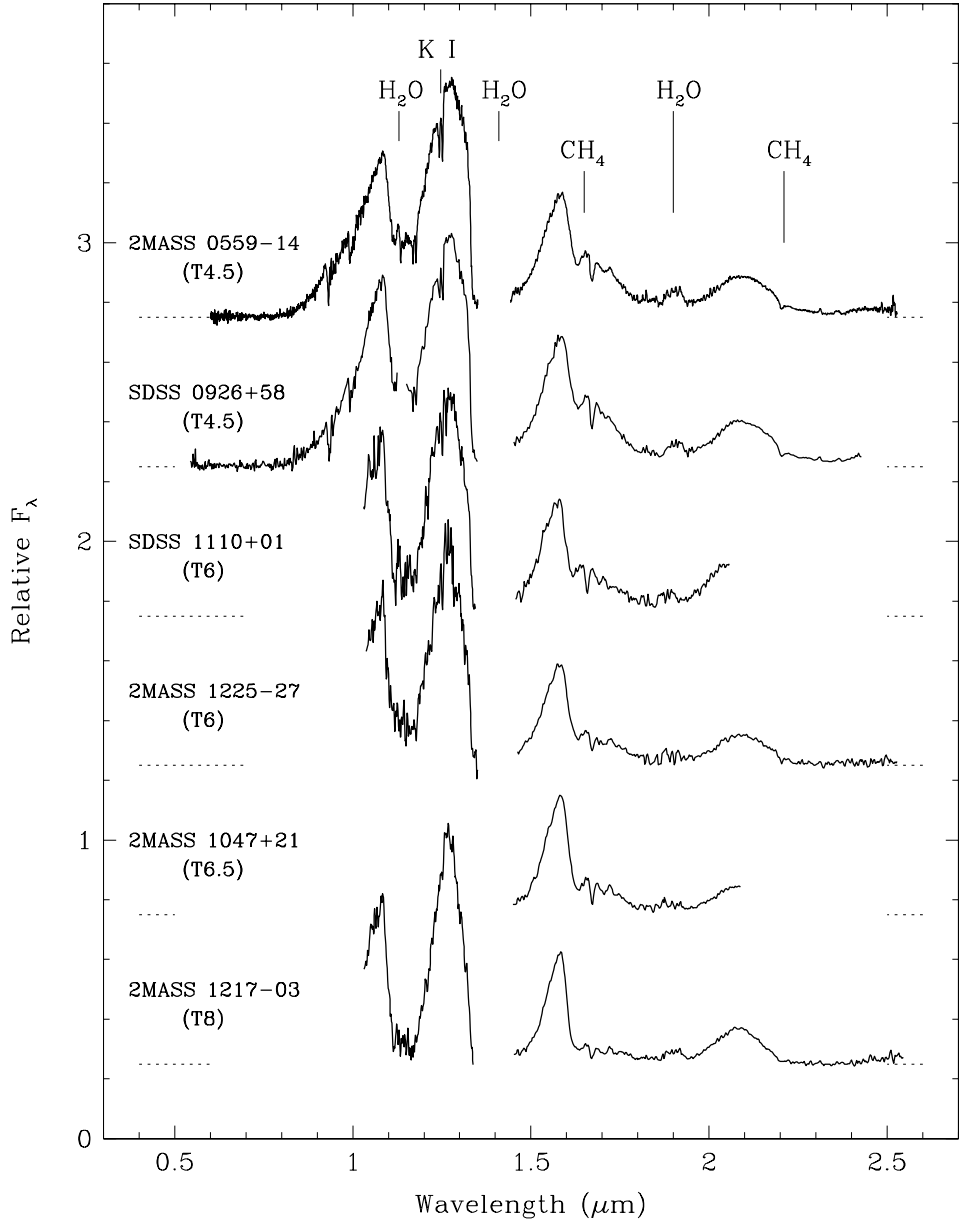}
\caption{T4.5 through T8.}
\end{figure}  

\newpage

\begin{figure}
\figurenum{4a}
\epsscale{.9}
\plotone{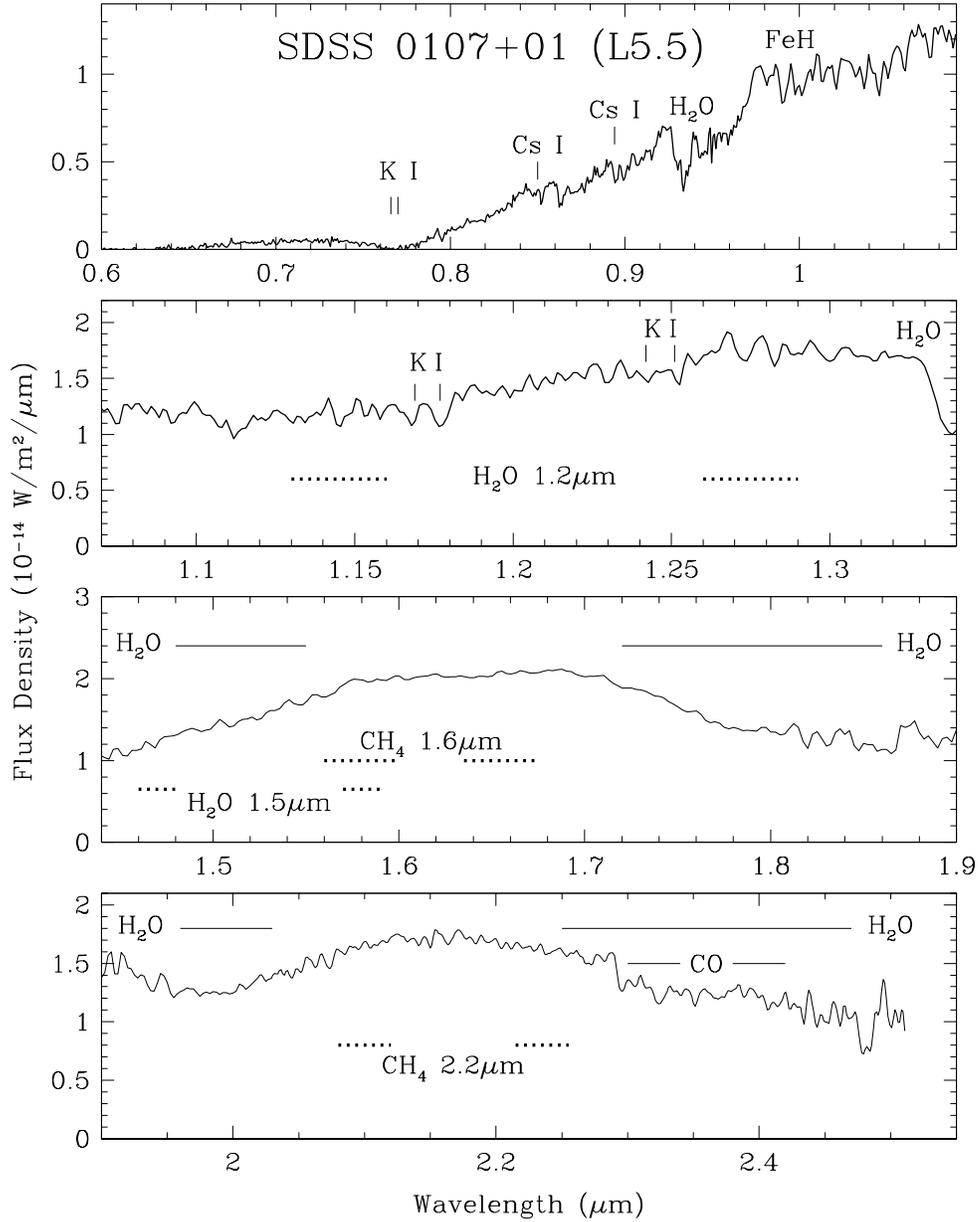}
\caption{Spectra of a mid-L dwarf and a mid-T dwarf
from 0.6 to 2.5~$\mu$m. The wavelengths of
prominent spectral lines and bands are marked, and the wavelength ranges
where the water and methane indices are calculated are indicated by
dotted lines. The L5.5 dwarf SDSS~0107+01.}
\end{figure}

\newpage

\begin{figure}
\figurenum{4b}
\epsscale{.9}
\plotone{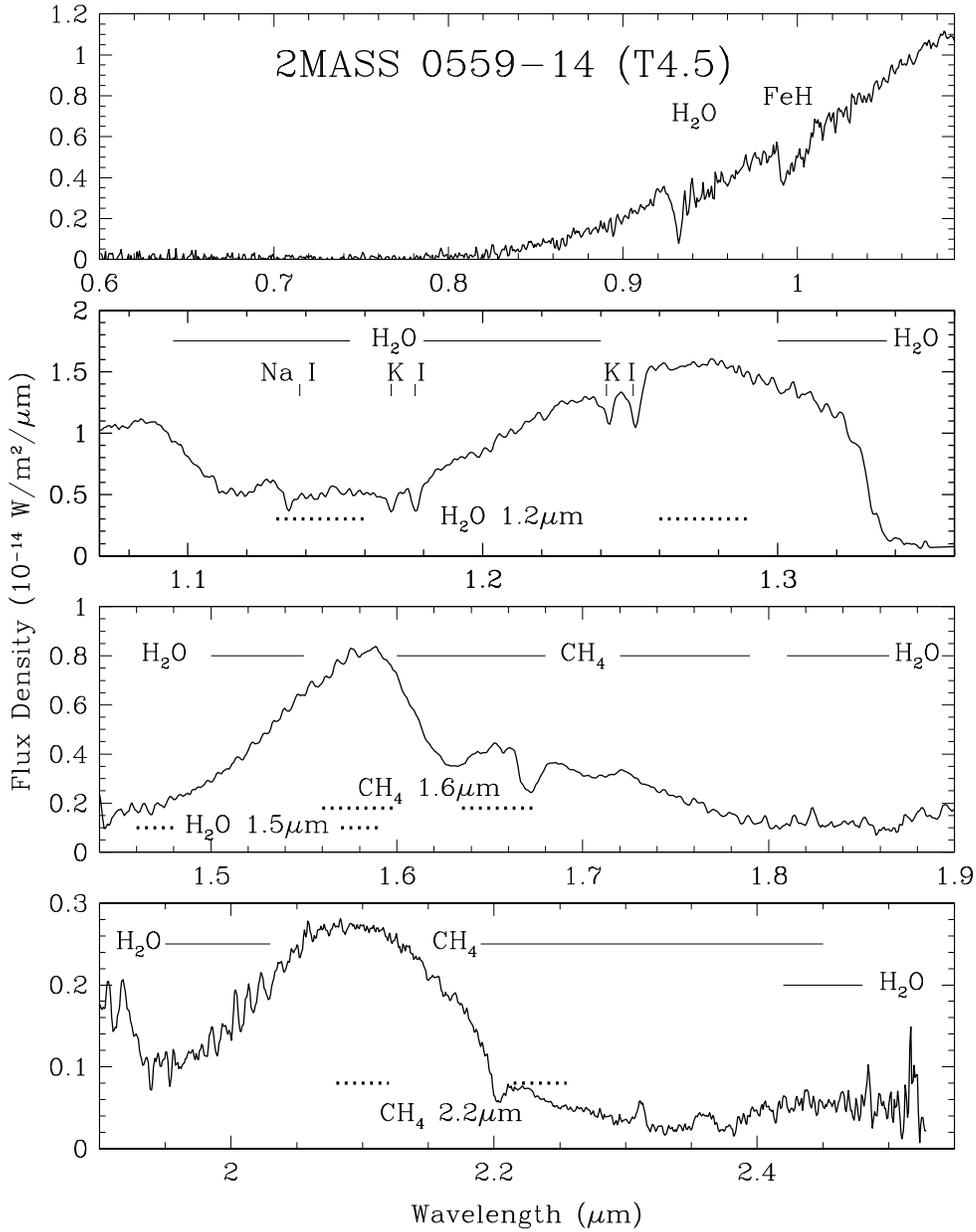} 
\caption{The T4.5 dwarf 2MASS~0559-14.}
\end{figure}

\newpage

\begin{figure}
\figurenum{5}
\epsscale{.9}
\plotone{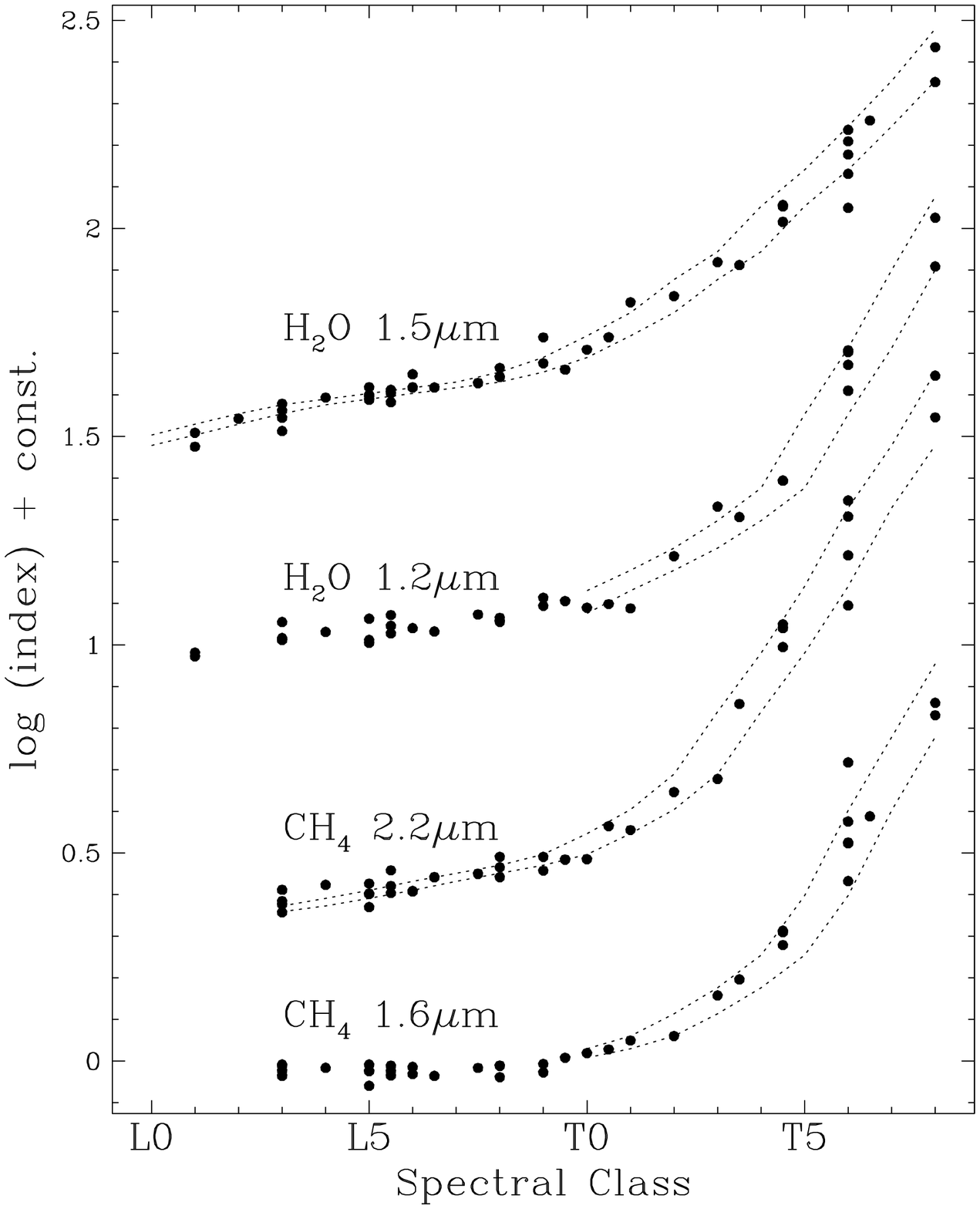}
\caption{Index values versus final assigned spectral class for the four
infrared indices. The space between the pairs of dashed lines denotes the
range of values of the index for each spectral class, according to
the definitions in Table~5.}
\end{figure}

\newpage

\begin{figure}
\figurenum{6a}
\epsscale{.9}
\plotone{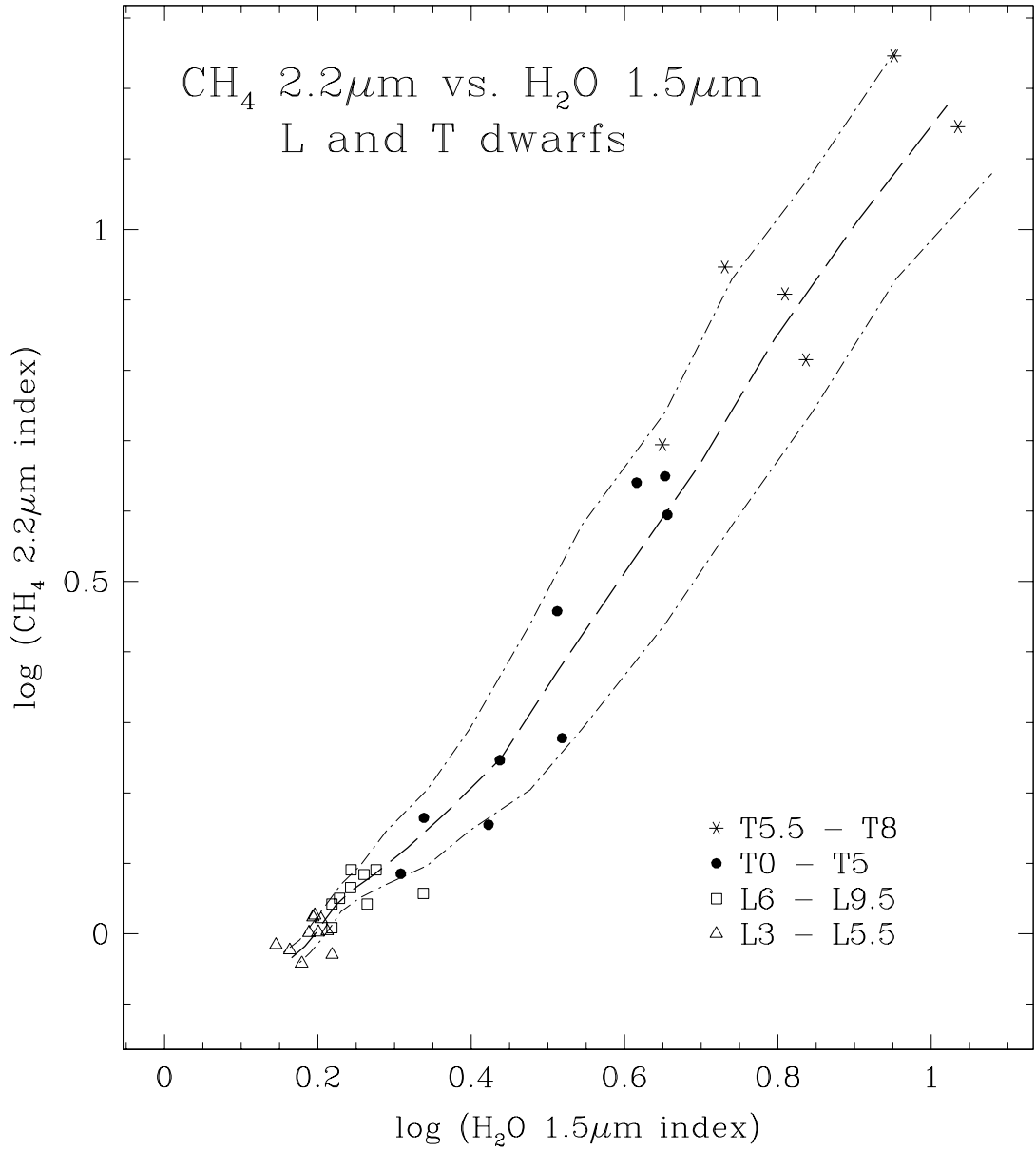}
\caption{The methane 2.2~$\mu$m index plotted against the water 1.5~$\mu$m
index, for L and T spectral classes. The dashed line connects the mean 
values of the subclasses according to the definitions in Table~5; the 
dot-dashed lines differ from the mean by $\pm$ 1 subclass. L and T
classes.}

\end{figure}

\newpage

\begin{figure}
\figurenum{6b}
\epsscale{.9}
\plotone{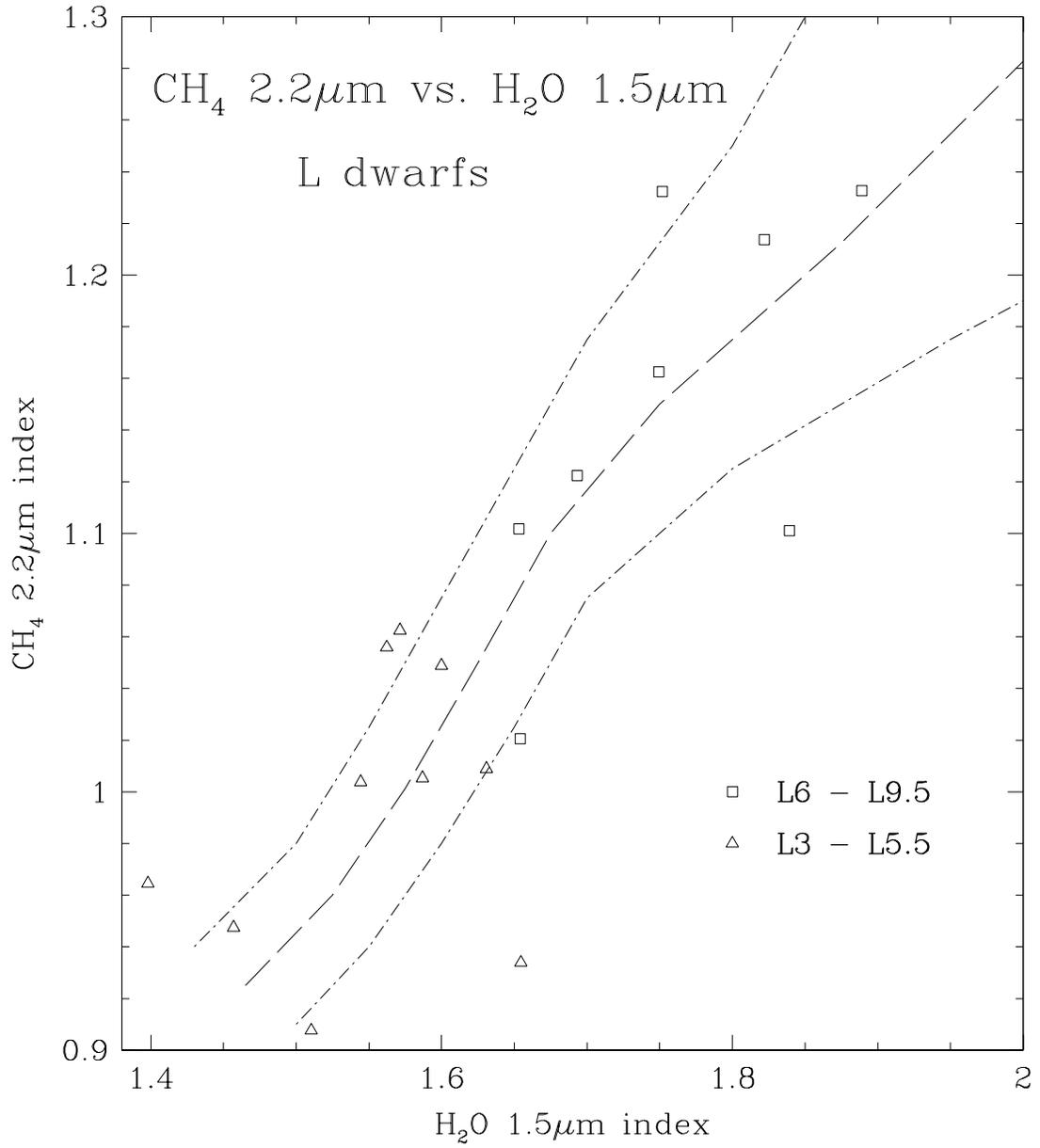}
\caption{L class only.}

\end{figure}

\begin{figure}
\figurenum{7}
\epsscale{.9}
\plotone{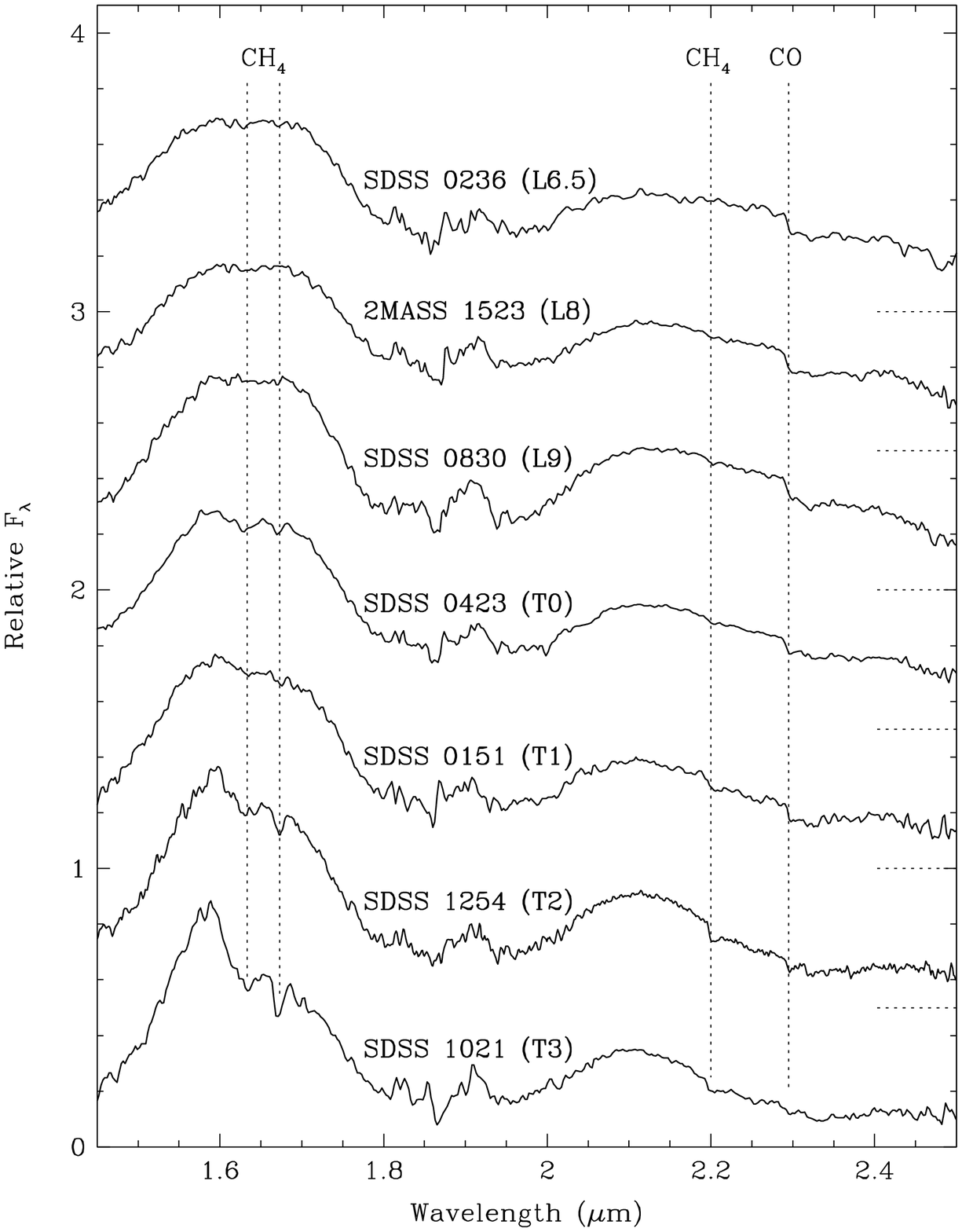} 
\caption{Spectra from 1.5 to 2.5~$\mu$m of late L and early T
dwarfs, showing the onset of absorption by methane in the H and K bands.
Horizontal dashed lines below each spectrum indicate zero flux levels.
Vertical dashed lines mark the wavelengths of spectral features due to
CH$_{4}$ and CO. Classifications are from this paper. Spectrum of
SDSS~1254-01 is from L00b.}
\end{figure}

\newpage

\begin{figure}
\figurenum{8}
\epsscale{.9}
\plotone{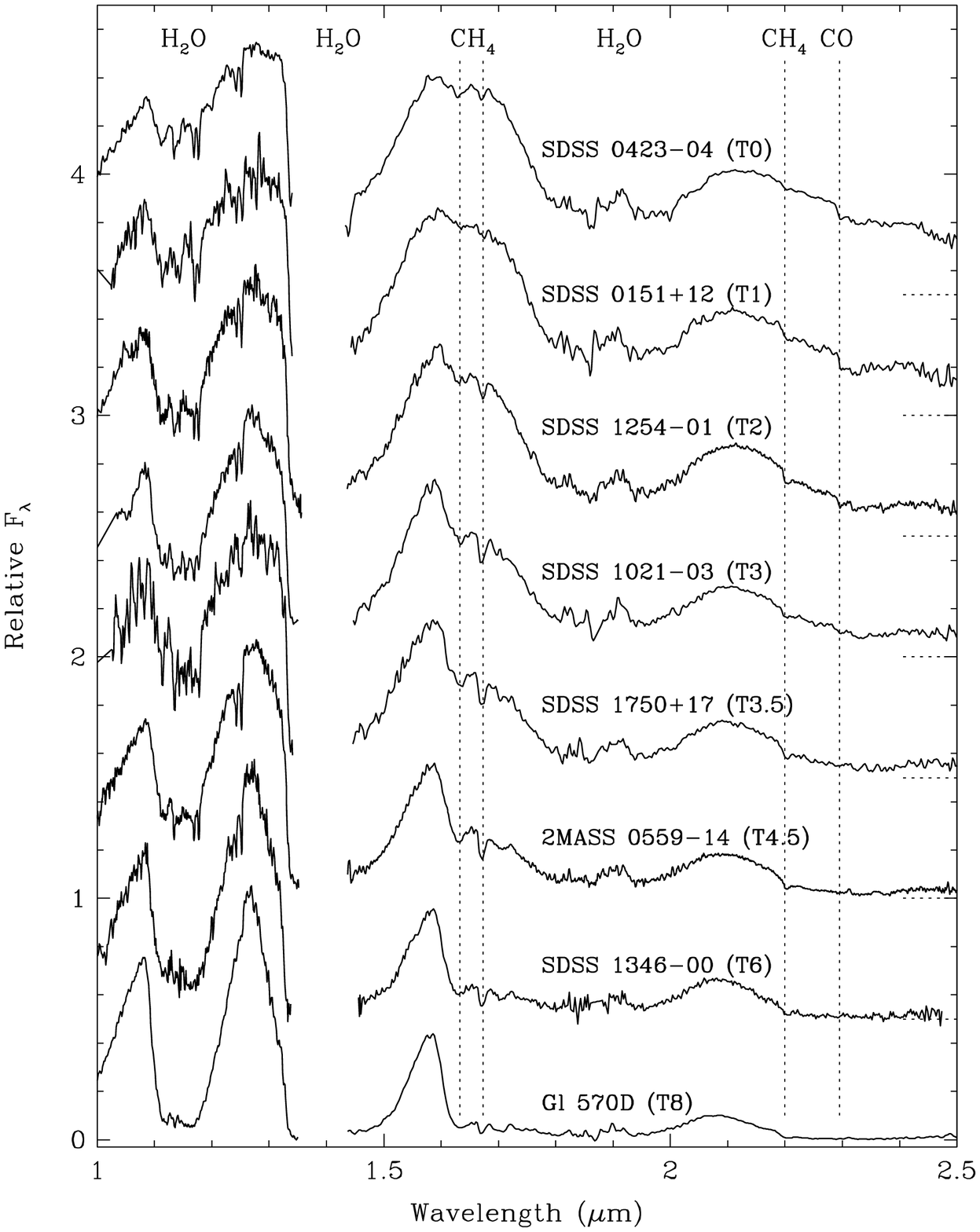}
\caption{The T sequence from 1.0 to 2.5~$\mu$m. The spectra are
normalized at their peaks near 1.3~$\mu$m. The wavelengths of broad water
absorption bands are indicated. Vertical dashed lines mark the wavelengths
of key spectral features due to CH$_{4}$ and CO. Classifications are from
this paper. Spectra of SDSS~1254-01, SDSS1346-00, and Gl~570D are from
L00b, Tsvetanov et al. (2000), and Geballe et al. (2001a), respectively.}
\end{figure}

\begin{figure}
\figurenum{9}
\epsscale{.9}
\plotone{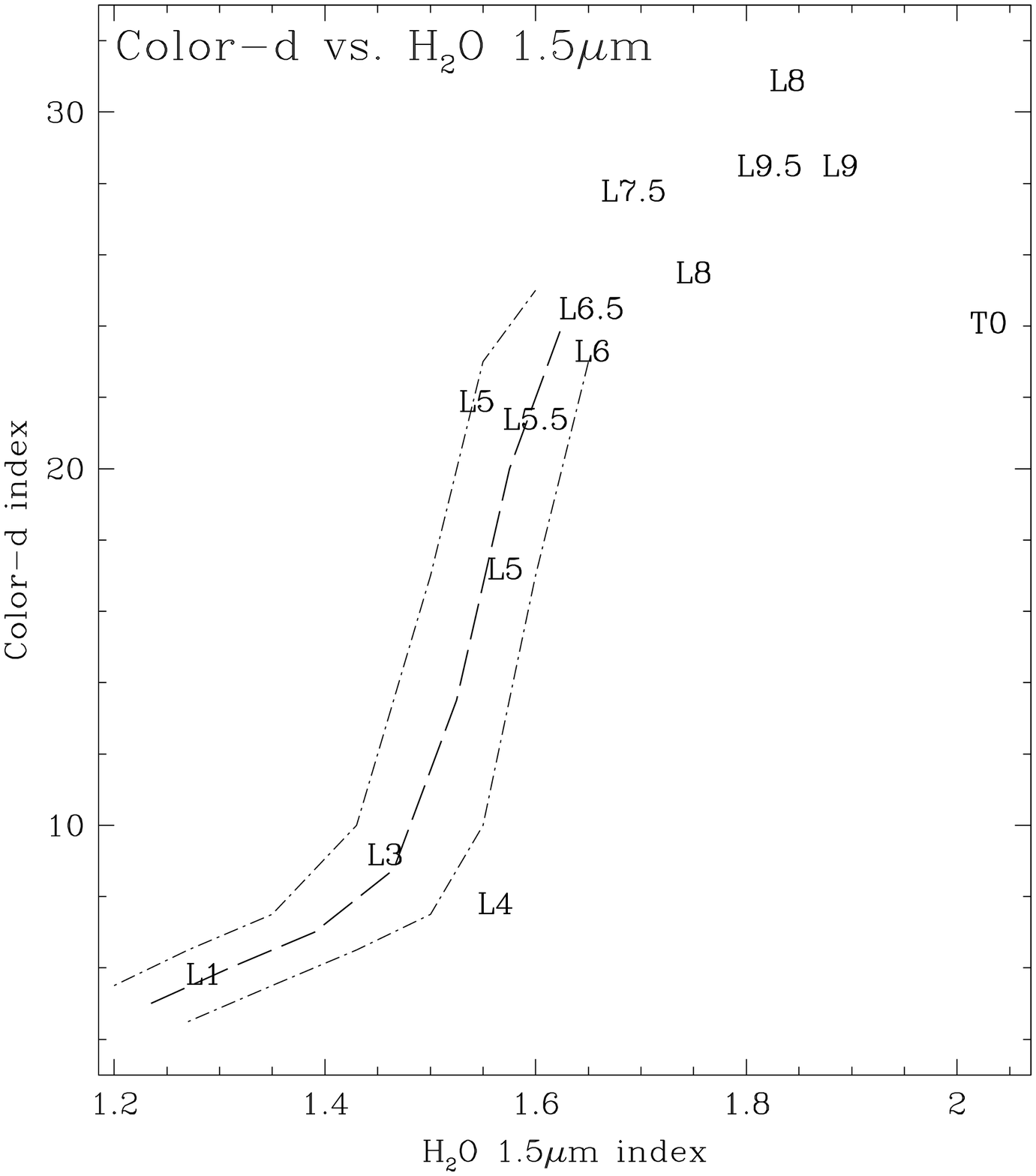}
\caption{The modified 2MASS optical ``Color-d'' index plotted against
the H$_{2}$O 1.5~$\mu$m index for isolated L dwarfs and the T0 dwarf
SDSS~0423-04. The dashed line connects the midpoints of the range of each
subclass where both are defined; the dash-dot lines deviate by one
subclass. Data points are from Table~4.}
\end{figure}

\newpage

\begin{deluxetable}{llllllll}
\scriptsize
\tablenum{1}
\tablewidth{0pt}
\tablecaption{Photometry of New SDSS Objects\tablenotemark{a,b,c}}
\tablehead{
\colhead{Name\tablenotemark{d}}&  \colhead{$r^{*}$}&
\colhead{$i^{*}$}&  \colhead{$z^{*}$}&
\colhead{J-K}& \colhead{J-H}&
\colhead{H-K}& \colhead{K} \nl
}
\tablecolumns{8}
\startdata
SDSSp J003259.36+141036.6& (24.18$\pm$0.56)& (23.02$\pm$0.37)& 19.52$\pm$0.08& ~1.59$\pm$0.07& ~0.92$\pm$0.07& ~0.67$\pm$0.07& 14.99$\pm$0.05\nl
SDSSp J010752.33+004156.1& (23.81$\pm$0.53)& 21.48$\pm$0.13& 18.72$\pm$0.04& ~2.17$\pm$0.04& ~1.19$\pm$0.04& ~0.98$\pm$0.04& 13.58$\pm$0.03\nl
SDSSp J015141.69+124429.6& (24.12$\pm$0.55)& (22.88$\pm$0.35)& 19.51$\pm$0.07& ~1.07$\pm$0.07& ~0.71$\pm$0.07& ~0.36$\pm$0.07& 15.18$\pm$0.05\nl
SDSSp J020742.83+000056.2& (23.90$\pm$0.49)&  22.55$\pm$0.26& 20.11$\pm$0.15& ~0.01$\pm$0.07& -0.03$\pm$0.07& ~0.04$\pm$0.07& 16.62$\pm$0.05\nl
SDSSp J023617.93+004855.0& (24.53$\pm$0.58)& 21.56$\pm$0.14& 18.94$\pm$0.05& ~1.47$\pm$0.07& ~0.85$\pm$0.07& ~0.62$\pm$0.07& 14.54$\pm$0.05\nl
SDSSp J042348.57$-$041403.5&  22.64$\pm$0.20& 20.22$\pm$0.04& 17.33$\pm$0.03& ~1.34$\pm$0.04& ~0.79$\pm$0.04& ~0.55$\pm$0.04& 12.96$\pm$0.03\nl
SDSSp J083008.12+482847.4& (23.59$\pm$0.48)& 21.26$\pm$0.08& 18.08$\pm$0.03& ~1.54$\pm$0.04& ~0.82$\pm$0.04& ~0.72$\pm$0.04& 13.68$\pm$0.03\nl
SDSSp J085758.45+570851.4&  23.02$\pm$0.30&  20.71$\pm$0.06& 17.73$\pm$0.02& ~1.86$\pm$0.04& ~1.00$\pm$0.04& ~0.86$\pm$0.04& 12.94$\pm$0.03\nl
SDSSp J092615.38+584720.9& (24.96$\pm$0.88)& (24.29$\pm$1.03)& 19.04$\pm$0.06& -0.03$\pm$0.04& ~0.05$\pm$0.04& -0.08$\pm$0.04& 15.50$\pm$0.03\nl
SDSSp J111010.01+011613.1&                 & (23.98$\pm$0.82)& 19.64$\pm$0.10& ~0.07$\pm$0.07& -0.10$\pm$0.07& ~0.17$\pm$0.07& 16.05$\pm$0.05\nl 
SDSSp J125737.26$-$011336.1&  22.71$\pm$0.27&  20.67$\pm$0.08& 18.52$\pm$0.05& ~1.58$\pm$0.04& ~0.96$\pm$0.04& ~0.62$\pm$0.04& 14.06$\pm$0.03\nl
SDSSp J131415.52$-$000848.1& (23.32$\pm$0.28)& 21.63$\pm$0.10& 19.63$\pm$0.07& ~1.05$\pm$0.07& ~0.49$\pm$0.07& ~0.56$\pm$0.07& 15.28$\pm$0.05\nl
SDSSp J144600.60+002452.0& (23.35$\pm$0.29)& 20.74$\pm$0.05& 18.54$\pm$0.05& ~1.76$\pm$0.07& ~0.97$\pm$0.07& ~0.79$\pm$0.07& 13.80$\pm$0.05\nl
SDSSp J175032.96+175903.9& (24.89$\pm$0.47)& (23.50$\pm$0.35)& 19.63$\pm$0.06&  ~0.12$\pm$0.07& ~0.20$\pm$0.07& -0.08$\pm$0.07& 16.02$\pm$0.05\nl
SDSSp J224953.45+004404.2& (23.88$\pm$0.57)& 22.05$\pm$0.24& 19.47$\pm$0.10& ~2.03$\pm$0.07& ~1.04$\pm$0.07 & ~0.99$\pm$0.07& 14.43$\pm$0.05\nl
SDSSp J225529.09$-$003433.4&  22.23$\pm$0.16&  19.86$\pm$0.03& 18.03$\pm$0.03& ~1.22$\pm$0.07& ~0.70$\pm$0.07& ~0.52$\pm$0.07& 14.28$\pm$0.05\nl
\enddata
\tablenotetext{a}{The SDSS photometry is reported in asinh magnitudes
(Lupton, Gunn, \& Szalay 1999).  Zero flux corresponds
to $z{^*}$=24.8,
$i^{*}$=24.4 and $z^{*}$=22.8. Values in parentheses are less than
5~$\sigma$ detections.}
\tablenotetext{b}{The SDSS magnitudes are AB$_{\nu}$ (Fukugita et al
1996) and the UKIRT magnitudes are on the Vega
system.}
\tablenotetext{c}{UKIRT (JHK) photometry is on the MKO system.}
\tablenotetext{d}{The `p' in SDSSp stands for "preliminary
astrometry." Coordinates are J2000.}
\end{deluxetable}

\newpage

%\hoffset=-.65truein
%\voffset=-.65truein
\begin{deluxetable}{lll}
\tablenum{2}
\tablewidth{0pt}
\tablecaption{Optical Spectroscopy}
\tablehead{  
\colhead{Name} &  \colhead{Telescope/} & \colhead{Date} \nl
\colhead{} & \colhead{Spectrograph} & \colhead{}}
\tablecolumns{3}
\startdata
SDSSp J010752.33+004156.1&  HET/LRS &  10/16/1999 \nl
SDSSp J015141.69+124429.6& ARC/DIS & 09/02/2000 \nl
SDSSp J023617.93+004855.0& HET/LRS & 10/06/1999 \nl
SDSSp J042348.57-041403.5& ARC/DIS & 01/19/2001 \nl
2MASSW J055919.1-140448&   ARC/DIS & 12/20/2000 \nl
SDSSp J083008.12+482847.4& ARC/DIS & 01/19/2001 \nl
SDSSp J085758.45+570851.4& ARC/DIS & 02/02/2001 \nl
SDSSp J092615.38+584720.9& ARC/DIS & 01/19/2001 \nl
SDSSp J102109.69-030420.1& ARC/DIS & 02/02/2001 \nl
SDSSp J144600.60+002452.0&  ARC/DIS & 03/01/2000 \nl
SDSSp J175032.96+175903.9& ARC/DIS & 02/02/2001 \nl
SDSSp 225529.09-003433.4&  HET/LRS & 12/07/1999 \nl
\enddata
\end{deluxetable}

\newpage

%\hoffset=-.65truein
%\voffset=-.65truein
\begin{deluxetable}{lllrl}
\scriptsize
\tablenum{3}
\tablewidth{0pt}
\tablecaption{Infrared Spectra: Observations}
\tablehead{
\colhead{Name} & \colhead{Telescope/}& \colhead{Bands}&
\colhead{Date} \nl
\colhead{} & \colhead{Spectrograph} & \colhead{} & \colhead{}
}
\tablecolumns{4}
\startdata
2MASSI J002839.44+150141.8 & UKIRT/CGS4 & J & 10/13/2000 \nl
 & UKIRT/CGS4 & H,K & 12/06/2000 \nl
 & Keck/NIRSPEC & $\rm J_s$ & 12/23/2000 \nl
SDSSp J003259.36+141036.6 &  UKIRT/CGS4 & Z,J,H,K & 09/20-23/2000 \nl
SDSSp J010752.33+004156.1 &  UKIRT/CGS4 & J,H,K & 09/20-23/2000 \nl
SDSSp J015141.69+124429.6 & UKIRT/CGS4 & J,H,K & 09/20-23/2000 \nl
SDSSp J020742.83+000056.2 & IRTF/NSFCAM & Z &  09/23/2000\nl
 & UKIRT/CGS4 & J& 12/07/2000 \nl
 & Keck/NIRSPEC & H,HK,K & 12/22-23/2000 \nl
SDSSp J023617.93+004855.0 & UKIRT/CGS4 & J,H,K & 09/20-23/2000 \nl
2MASSW J031059.9+164816 & UKIRT/CGS4 & J & 07/12/2001 \nl
                        & UKIRT/CGS4 & H & 07/18/2001 \nl
2MASSW J032842.6+230205 & UKIRT/CGS4 & J,H,K & 12/05-07/2000 \nl
SDSSp J042348.57-041403.5 & UKIRT/CGS4 & Z,J,H,K & 09/20-23/2000 \nl
2MASSW J055919.1-140448 & UKIRT/CGS4 & J,H,K & 09/20-23/2000 \nl
 & UKIRT/CGS4 & Z,J & 12/05-07/2000 \nl
SDSSp J083008.12+482847.4 & UKIRT/CGS4 & J,H,K & 12/05-07/2000 \nl
SDSSp J085758.45+570851.4 &UKIRT/CGS4 & J,H,K & 12/05-07/2000 \nl
SDSSp J092615.38+584720.9 & Keck/NIRSPEC & $\rm J_s$,J,H,HK,K & 12/21-23/2000 \nl
SDSSp J102109.69-030420.1 & Keck/NIRSPEC & $\rm J_s$ & 12/22/2000 \nl
2MASSI J104753.9+212423   & UKIRT/CGS4 & H & 05/21/2001 \nl
SDSSp J111010.01+011613.1 & UKIRT/CGS4 & H & 05/21/2001 \nl
 & UKIRT/CGS4 & J & 5/24/2001 \nl
2MASSW J121711.1-031113   & UKIRT/CGS4 & H & 05/21/2001 \nl
 & UKIRT/CGS4 & J,K & 5/24/2001 \nl  
2MASSW J1225543-273947    & UKIRT/CGS4   & H        & 06/16/2001 \nl
 & UKIRT/CGS4 & J & 07/12/2001 \nl
 & UKIRT/CGS4 & K & 07/18/2001 \nl                  
SDSSp J125737.26-011336.1 & Keck/NIRSPEC & J,H,HK,K & 12/22-23/2000 \nl
SDSSp J131415.52-000848.1 & Keck/NIRSPEC & H,K & 12/23/2000 \nl
SDSSp J132629.82-003831.5 & UKIRT/CGS4 & J,H,K & 03/13-15/2000 \nl
SDSSp J144600.60+002452.0 & UKIRT/CGS4 & J,H,K & 03/13-15/2000 \nl
2MASSW J152322.6+301456 & UKIRT/CGS4 & J,H,K & 03/13-15/2000 \nl
2MASSW J163229.1+190441 & UKIRT/CGS4 & Z,J,H,K & 09/20-23/2000 \nl
SDSSp J175032.96+175903.9 & UKIRT/CGS4 & J,H & 02/19/2001 \nl
 & UKIRT/CGS4 & K & 05/06/2001 \nl
SDSSp J224953.45+004404.2 & UKIRT/CGS4 & J,H,K & 12/05-07/2000 \nl
 &  Keck/NIRSPEC & $\rm J_s$ & 12/23/2000 \nl
SDSSp J225529.09-003433.4 & UKIRT/CGS4 & J,H,K & 09/20-23/2000 \nl
\enddata\end{deluxetable}

\newpage
%\hoffset=-.65truein
%\voffset=-.65truein

\begin{deluxetable}{llllllllll}   
\tiny
\tablenum{4}
\tablewidth{0pt}
\tablecaption{Spectral Indices\tablenotemark{a}~~and Classifications of Cool Dwarfs}
\tablehead{
\colhead{Name} & \colhead{PC3\tablenotemark{b}} & \colhead{Color-d\tablenotemark{c}} & \colhead{Cont 1.0$\mu$m} & \colhead{H$_{2}$O
1.2$\mu$m} & \colhead{H$_{2}$O 1.5$\mu$m} & \colhead{CH$_{4}$ 1.6$\mu$m} & \colhead{H$_{2}$O 2.0$\mu$m} & \colhead{CH$_{4}$
2.2$\mu$m} & \colhead{Assigned} \nl   
\colhead{} & \colhead{0.823--0.827/} & \colhead{0.96--0.98/} & \colhead{1.04--1.05/} & \colhead{1.26--1.29/} 
& \colhead{1.57--1.59/} & \colhead{1.56--1.60/} & \colhead{2.09--2.11/} &\colhead{2.08--2.12/} & \colhead{Class\tablenotemark{d}} \nl
\colhead{} & \colhead{0.754--0.758} & \colhead{0.735--0.755} & \colhead{0.875--0.885} & \colhead{1.13--1.16} &
\colhead{1.46--1.48} & \colhead{1.635--1.675} & \colhead{1.975--1.995} & \colhead{2.215--2.255} &}
\tablecolumns{10}
\startdata
LHS 65         & \nodata    & 0.85        & \nodata & 0.82    & \nodata      & \nodata  & 0.91       & \nodata     & M1       \nl
LHS 38         & 0.96       & 0.47        & 0.97    & 0.84    & \nodata      & \nodata  & 0.96       & \nodata     & M1       \nl
LHS 386        & 1.03       & 0.93        & 0.42    & 0.87    & \nodata      & \nodata  & 0.94       & \nodata     & M1       \nl
LHS 443        & \nodata    & \nodata     & \nodata & \nodata & 0.99         & \nodata  & 0.94       & \nodata     & M1.5     \nl
LHS 399        & \nodata    & \nodata     & \nodata & \nodata & 0.98         & \nodata  & 0.97       & \nodata     & M2.5     \nl
LHS 502        & \nodata    & \nodata     & \nodata & \nodata & 0.94         & \nodata  & 0.97       & \nodata     & M2.5     \nl
LHS 5327       & \nodata    & \nodata     & 0.81    & 0.93    & 0.91         & \nodata  & 0.96       & \nodata     & M3       \nl
LHS 54         & \nodata    & \nodata     & \nodata & \nodata & 1.03         & \nodata  & 0.96       & \nodata     & M3       \nl
LHS 58         & 1.00       & \nodata     & 0.95    & 0.87    & 0.90         & \nodata  & 0.92       & \nodata     & M3       \nl
LHS 2945       & 1.11       & 1.21        & 1.07    & 0.93    & 0.94         & \nodata  & 0.97       & \nodata     & M3.5     \nl
LHS 4          & 0.96       & 1.15        & 0.95    & 1.01    & 1.00         & \nodata  & 0.98       & \nodata     & M3.5     \nl
LHS 427        & 1.00       & \nodata     & \nodata & \nodata & 1.06         & \nodata  & 0.96       & \nodata     & M3.5     \nl
LHS 59         & 1.12       & \nodata     & \nodata & \nodata & 0.97         & \nodata  & 0.94       & \nodata     & M3.5     \nl
LHS 57         & 1.12       & 1.35        & 0.92    & 0.95    & \nodata      & \nodata  & 0.99       & \nodata     & M4       \nl
LHS 3509       & \nodata    & \nodata     & \nodata & \nodata & 0.97         & \nodata  & 1.01       & \nodata     & M4.5     \nl
LHS 400        & \nodata    & \nodata     & \nodata & \nodata & 1.03         & \nodata  & 1.04       & \nodata     & M4.5     \nl
LHS 421        & 1.17       & 1.24        & 0.92    & 0.91    & 1.08         & \nodata  & 0.95       & \nodata     & M4.5     \nl
LHS 3494       & 1.19       & 1.87        & 1.27    & 0.98    & 1.04         & \nodata  & 1.05       & \nodata     & M5.5     \nl
LHS 2          & \nodata    & 2.20        & 1.40    & 0.92    & \nodata      & \nodata  & 0.99       & \nodata     & M5.5     \nl
LHS 3406       & \nodata    & \nodata     & 1.68    & 0.94    & 1.08         & \nodata  & 1.08       & \nodata     & M5.5     \nl
LHS 39         & 1.32       & 1.81        & 1.35    & 1.05    & 1.04         & \nodata  & 1.04       & \nodata     & M5.5     \nl
LHS 330        & 1.42       & \nodata     & \nodata & \nodata & 1.08         & \nodata  & 1.03       &\nodata      & M6       \nl
LHS 3339       & 1.42       & \nodata     & \nodata & 1.04    & 1.02         & \nodata  & 1.00       & \nodata     & M6       \nl
LHS 3495       & 1.37       & \nodata     & \nodata & 0.92    & 1.01         & \nodata  & 1.01       & \nodata     & M6       \nl
LHS 36         & 1.62       & 2.5         & \nodata & 1.00    & 1.03         & \nodata  & 1.04       & \nodata     & M6       \nl
LHS 5328       & \nodata    & \nodata     & 1.26    & 1.03    & 1.02         & \nodata  & 1.04       & \nodata     & M6       \nl
LHS 292        & 1.58       & \nodata     & \nodata & 0.97    & \nodata      & \nodata  & 1.04       & \nodata     & M6.5     \nl
LHS 2930       & 1.68       & 2.9         & 1.85    & 1.00    & 1.16         & \nodata  & 1.07       & \nodata     & M6.5     \nl
LHS 523        & 1.58       & \nodata     & 1.45    & 0.96    & 1.01         & \nodata  & 1.04       & \nodata     & M6.5     \nl
LHS 429        & 1.51       & 2.4         & 1.61    & 1.03    & \nodata      & \nodata  & 1.06       & \nodata     & M7       \nl
LHS 3003       & 1.67       & \nodata     & \nodata & 1.12    & 1.17         & \nodata  & 1.09       & \nodata     & M7       \nl
T 513          & 1.99       & \nodata     & 2.8     & 1.09    & 1.05         & \nodata  & 1.09       & \nodata     & M8.5     \nl
SDSS 2255-00   & \nodata    & \nodata     & 2.2     & 1.00    & 1.15         & \nodata  & 1.12       & \nodata     & M8.5     \nl
LHS 2065       & 1.95       & \nodata     & \nodata & 1.15    & 1.15         & \nodata  & 1.11       & \nodata     & M9       \nl
LP 944         & 2.5        & \nodata     & \nodata & 1.19    & 1.25         & \nodata  & 1.13       & \nodata     & M9       \nl
BRI 0021       & 2.3        & 4.7         & \nodata & 1.22    & 1.24         & \nodata  & 1.15       & \nodata     & M9.5     \nl
2MASS 0345+25  & 3.2 (L2.5) & \nodata     & \nodata & 1.20    & 1.19 ($<$L0) & \nodata  & 1.09       & \nodata     & L1$\pm$1 \nl
2MASS 0746+20  & 2.7 (L1)   & 5.8 (L1)    & 2.2     & 1.18    & 1.29 (L1)    & \nodata  & 1.21       & \nodata     & L1       \nl
SDSS 1314-00   & \nodata    & \nodata     & \nodata & \nodata & 1.39 (L2)    & \nodata  & \nodata    & \nodata     & L2(?)    \nl
Kelu-1         & 2.9 (L2.5) & \nodata     & 4.1     & 1.31    & 1.40 (L2)    & 0.98     & 1.27       & 0.96 (L4)   & L3$\pm$1 \nl
GD 165B        & 3.5 (L3)   & 8.8 (L3)    & 3.9     & 1.30    & 1.30 (L1)    & 0.98     & 1.36       & 1.03 (L5.5) & L3$\pm$2 \nl
DENIS 1058-15  & 3.5 (L3)   & 7.4 (L2.5)  & 3.5     & 1.29    & 1.51 (L3.5)  & 0.95     & \nodata    & 0.91 (L3)   & L3       \nl
2MASS 0028+15  & 4.0 (L3)   & 9.2 (L3)    & 3.4     & 1.43    & 1.46 (L3)    & 0.92     & 1.36       & 0.95 (L4)   & L3       \nl
2MASS 0036+18  & 3.8 (L3)   & 7.8 (L2.5)  & 3.4     & 1.35    & 1.56 (L4.5)  & 0.96     & 1.29       & 1.06 (L5)   & L4$\pm$1 \nl
SDSS 2249+00   & \nodata    & \nodata     & \nodata & 1.45    & 1.65 (L6.5)  & 0.87     & 1.29       & 0.93 (L3)   & L5$\pm$1 \nl   
SDSS 1446+00   & 7.0        & 22. (L5.5)  & 3.1     & 1.27    & 1.54 (L4.5)  & 0.95     & 1.39       & 1.01 (L5)   & L5       \nl
SDSS 1257-01   & \nodata    & \nodata     & \nodata & \nodata & 1.59 (L5.5)  & 0.95     & 1.44       & 1.01 (L5)   & L5       \nl
SDSS 0539-00   & 8.5        & 17.2 (L4.5) & 3.0     & 1.29    & 1.57 (L5)    & 0.98     & 1.43       & 1.06 (L6)   & L5       \nl
SDSS 0107+00   & 6.4        & 21. (L5)    & 2.2     & 1.48    & 1.60 (L5.5)  & 0.95     & 1.41       & 1.05 (L6)   & L5.5     \nl
SDSS 1326-00   & \nodata    & \nodata     & \nodata & 1.40    & 1.63 (L6)    & 0.92     & 1.26       & 1.01 (L5)   & L5.5     \nl
DENIS 0205-11  & 11.1       & 15.2 (L4)   & 3.3     & 1.34    & 1.52 (L4)    & 0.97     & 1.51       & 1.14 (L8) & L5.5$\pm$2 \nl
DENIS 1228-15  & 6.7        & 14.1 (L4)   & 4.0     & 1.38    & 1.78 (L8.5)  & 0.97     & 1.39       & 1.01 (L5)   & L6$\pm$2 \nl
2MASS 0825+21  & 10.8       & 23. (L5.5)  & 3.2     & \nodata & 1.65 (L6.5)  & 0.93     & 1.44       & 1.02 (L5.5) & L6       \nl
SDSS 0236+00   & 13.5       & 25. (L6.5)  & 3.2     & 1.35    & 1.65 (L6.5)  & 0.92     & 1.49       & 1.10 (L7)   & L6.5     \nl
2MASS 1632+19  & 12.3       & 28.         & 2.5     & 1.49    & 1.69 (L7.5)  & 0.96     & 1.45       & 1.12 (L7.5) & L7.5     \nl
2MASS 1523+30  & 11.7       & 25. (L6.5)  & 2.9     & 1.43    & 1.75 (L8)    & 0.97     & 1.42       & 1.16 (L8)   & L8       \nl
SDSS 0857+57   & 17.5       & 31.         & 2.7     & 1.44    & 1.84 (L9)    & 0.97     & 1.49       & 1.10 (L7)   & L8$\pm$1 \nl  
SDSS 0032+14   & \nodata    & \nodata  & \nodata & 1.46      & 1.75 (L8)   & 0.92       & 1.48       & 1.23 (L8.5) & L8       \nl
SDSS 0830+48   & \nodata    & \nodata  & \nodata & 1.56 (T0) & 2.18 (T0.5) & 0.98 ($<$T0) & 1.91     & 1.14 (L8)   & L9$\pm$1 \nl
2MASS 0310+16  & 8.1        & 29.      & 2.8     & 1.63 (T0) & 1.89 (L9)   & 0.94 ($<$T0) & 1.19     & 1.23 (L9)   & L9       \nl
2MASS 0328+23  & 9.6        & 29.      & 2.5     & 1.60 (T0) & 1.82 (L8.5) & 1.02 (L9.5)  & 1.55     & 1.21 (L9) & L9.5$\pm$1 \nl
SDSS 0423-04   & 7.1        & 24.      & 3.0     & 1.54 (T0) & 2.03 (T0)   & 1.05 (T0)    & 1.51     & 1.22 (L9)   & T0       \nl
SDSS 0837-00   & \nodata    & 19.8     & 4.5     & 1.58 (T0) & 2.18 (T0.5) & 1.07 (T0.5)  & 1.67     & 1.46 (T1)   & T0.5     \nl
SDSS 0151+12   & 7.8        & 28.      & 3.1     & 1.54 (L9.5) & 2.65 (T2) & 1.12 (T1)    & 1.53     & 1.43 (T0.5) & T1$\pm$1 \nl
SDSS 1254-01   & \nodata    & \nodata  & 6.5     & 2.06 (T2)   & 2.74 (T2) & 1.15 (T1.5)  & 1.83     & 1.76 (T2)   & T2       \nl
SDSS 1021-03   & \nodata    & \nodata  & 4.3     & 2.70 (T4) & 3.3 (T3)    & 1.44 (T3)    & 1.78     & 1.90 (T2.5) & T3       \nl
SDSS 1750+17   & \nodata    & \nodata  & 4.7     & 2.55 (T3.5) & 3.3 (T3)  & 1.57 (T4)    & 2.00     & 2.87 (T3.5) & T3.5     \nl
SDSS 0926+58   & \nodata    & \nodata  & 7.9     & \nodata   & 4.5 (T4.5)  & 2.04 (T5)    & 2.13     & 3.9 (T4.5)  & T4.5     \nl
SDSS 0207+00   & \nodata    & \nodata  & \nodata & 3.1 (T4.5) & 4.5 (T4.5) & 1.90 (T4.5)  & 2.08     & 4.5 (T5)    & T4.5     \nl
2MASS 0559-14  & 3.5        & 88.      & 6.7     & 3.1 (T4.5) & 4.1 (T4)   & 2.06 (T5)    & 2.11     & 4.4 (T5)    & T4.5     \nl
Gl 229B        & \nodata    & \nodata  & 15.5    & 6.4 (T6.5) & 4.6 (T4.5) & 5.2 (T7)     & 2.45     & 4.9 (T5)    & T6$\pm$1 \nl
SDSS 1110+01   & \nodata    & \nodata  & \nodata & 5.1 (T6)  & 6.0 (T6)    & 2.71 (T6)    & \nodata  & \nodata     & T6       \nl
2MASS 1225-27  & \nodata    & \nodata  & \nodata & 5.0 (T6)  & 6.4 (T6)    & 3.3 (T6)     & 2.50     & 8.1 (T6)    & T6       \nl
SDSS 1346-00   & \nodata    & \nodata  & 6.1     & 5.9 (T6)  & 5.4 (T5.5)  & 3.8 (T6)     & 2.38     & 8.8 (T6.5)  & T6       \nl
SDSS 1624+00   & 15.1       & 83.      & 7.7     & 6.4 (T6.5) & 6.9 (T6.5) & 3.4 (T6)     & 2.17     & 6.5 (T6)    & T6       \nl
2MASS 1047+21  & \nodata    & \nodata  & \nodata & \nodata   & 7.2 (T6.5)  & 3.9 (T6.5)   & \nodata  & \nodata     & T6.5     \nl
2MASS 1217-03  & \nodata    & \nodata  & \nodata & 10.2 (T7.5) & 8.9 (T7.5) & 6.8 (T8)    & 2.71     & 17.6 (T8.5) & T8       \nl
Gl 570D        & \nodata    & \nodata  & 21.     & 13.3 (T8) & 10.8 (T8)   & 7.2 (T8)     & 2.29     & 14.0 (T8)   & T8 
\enddata
\tablenotetext{a}{Indices are defined in Table~5 and discussed in Section 4.2.}
\tablenotetext{b}{Index defined by M99 and discussed in Section 4.2.}
\tablenotetext{c}{Index modified from the original definition (K99), as discussed in Section 4.2.}
\tablenotetext{d}{Uncertainties are typically $\pm$0.5, except where noted.} 
\end{deluxetable}

\clearpage

%\hoffset=-.65truein
%\voffset=-.65truein
\begin{deluxetable}{lllllll}
\small
\tablenum{5}
%\tablewidth{400pt}
\tablecaption{Values of Indices\tablenotemark{a}~~for L and T Subtyping}
\tablehead{
\colhead{Name} & \colhead{PC3\tablenotemark{b}} & \colhead{Color-d\tablenotemark{c}} & \colhead{H$_{2}$O 1.2$\mu$m} &
\colhead{H$_{2}$O 1.5$\mu$m} & \colhead{CH$_{4}$ 1.6$\mu$m} &  \colhead{CH$_{4}$ 2.2$\mu$m} \nl
\colhead{} & \colhead{[0.823--0.827]/} & \colhead{[0.96--0.98]/} &
\colhead{[1.26--1.29]/} & \colhead{[1.57--1.59]/} & \colhead{[1.56--1.60]/} & \colhead{[2.08--2.12]/} \nl
\colhead{} & \colhead{[0.754--0.758]} & \colhead{[0.735--0.755]} & \colhead{[1.13--1.16]} & \colhead{[1.46--1.48]}
& \colhead{[1.635-1.675]} & \colhead{[2.215--2.255]}
}
\tablecolumns{7}
\small
\startdata
L0 & 2.4-2.6   & 4.5-5.5  &            & 1.20-1.27  &            &               \nl
L1 & 2.6-2.85  & 5.5-6.5  &            & 1.27-1.35  &            &               \nl
L2 & 2.85-3.25 & 6.5-7.5  &            & 1.35-1.43  &            &               \nl
L3 & 3.25-4.25 & 7.5-10.  &            & 1.43-1.50  &            & 0.91-0.94     \nl
L4 & 4.25-6.0  & 10-17    &            & 1.50-1.55  &            & 0.94-0.98     \nl
L5 &           & 17-23    &            & 1.55-1.60  &            & 0.98-1.025    \nl
L6 &           & 23-25    &            & 1.60-1.65  &            & 1.025-1.075   \nl
L7 &           &          &            & 1.65-1.70  &            & 1.075-1.125   \nl
L8 &           &          &            & 1.70-1.80  &            & 1.125-1.175   \nl
L9 &           &          &            & 1.80-1.95  &            & 1.175-1.25    \nl
T0 &           &          & 1.5-1.7    & 1.95-2.2   & 1.02-1.07  & 1.25-1.40     \nl
T1 &           &          & 1.7-1.9    & 2.2-2.5    & 1.07-1.15  & 1.40-1.60     \nl
T2 &           &          & 1.9-2.15   & 2.5-3.0    & 1.15-1.30  & 1.60-1.95     \nl
T3 &           &          & 2.15-2.5   & 3.0-3.5    & 1.30-1.50  & 1.95-2.75     \nl
T4 &           &          & 2.5-3.0    & 3.5-4.5    & 1.50-1.80  & 2.75-3.8      \nl
T5 &           &          & 3.0-4.5    & 4.5-5.5    & 1.80-2.50  & 3.8-5.5       \nl
T6 &           &          & 4.5-6.5    & 5.5-7.0    & 2.5-4.0    & 5.5-8.5       \nl
T7 &           &          & 6.5-10.    & 7.0-9.0    & 4.0-6.0    & 8.5-12.       \nl
T8 &           &          & 10.-15.(?) & 9.0-12.(?) & 6.0-9.0(?) & 12.-18.(?)    \nl
\enddata
\tablenotetext{a}{Indices are discussed in Section 4.2.}
\tablenotetext{b}{Index defined by M99 and discussed in Section 4.2.}
\tablenotetext{c}{Index modified from the original definition (K99), as discussed in Section 4.2.}
\end{deluxetable}
\end{document}